\documentclass[12pt]{article}
\usepackage[utf8]{inputenc}
\usepackage{amsmath}
\usepackage{amsthm}
\usepackage{mathtools}
\usepackage{amssymb}
\usepackage{commath}
\usepackage{setspace}
\usepackage{changepage}
\usepackage{tikz}
\usepackage{setspace}
\usepackage{multirow}
\usepackage{natbib}
\usepackage{booktabs}
\usepackage{setspace,float}
\usepackage{footmisc}
\usepackage{bbm}
\usepackage{threeparttable}
\usepackage{charter}
\usepackage{xfrac}
\usepackage{xcolor}
\usepackage{authblk}
\usepackage{mathrsfs}
\usepackage[english]{babel}
\usepackage[ruled,vlined]{algorithm2e}
\usepackage{tikz}
\usepackage[title]{appendix}
\usepackage[font=small,labelfont=bf]{caption}
\usepackage{enumerate}
\usepackage[colorlinks,
urlcolor=blue,
linkcolor=blue, 
anchorcolor=blue,
citecolor=blue]{hyperref}
\usetikzlibrary{decorations.pathreplacing, arrows,positioning}
\renewcommand{\arraystretch}{1.5}
\newtheorem{theorem}{Theorem}[section]
\newtheorem{assumption}{Assumption}[section]

\newtheorem{lemma}{Lemma}[section]
\newtheorem{proposition}{Proposition}[section]
\theoremstyle{definition}
\newtheorem{definition}{Definition}[section]
\newtheorem{remark}{Remark}[section]
\newcommand*\circled[1]{\tikz[baseline=(char.base)]{
            \node[shape=circle,draw,inner sep=1pt] (char) {#1};}}
\DeclareMathOperator{\E}{\mathbb{E}}

\providecommand{\keywords}[1]
{
  \small	
  \textbf{\textit{Keywords:}} #1
}
\DefineFNsymbols{mySymbols}{{\ensuremath\dagger}{\ensuremath\ddagger}\S\P
   *{**}{\ensuremath{\dagger\dagger}}{\ensuremath{\ddagger\ddagger}}}
\setfnsymbol{mySymbols}

\usepackage[a4paper]{geometry}
\begin{document}
\linespread{1.25}
\title{\vspace{-1cm} Who Should Get Vaccinated? Individualized Allocation of Vaccines Over SIR Network}
\author{Toru Kitagawa\thanks{Department of Economics, University College London. Email$\colon$ \href{mailto:t.kitagawa@ucl.ac.uk}{t.kitagawa@ucl.ac.uk}} and Guanyi Wang\thanks{Department of Economics, University College London. Email$\colon$ \href{mailto:guanyi.wang.17@ucl.ac.uk}{guanyi.wang.17@ucl.ac.uk}\\
We would like to thank Andrew Chesher, Raffaella Giacomini, Aureo de Paula, Martin Weidner and the participants at 2020 UCL Economics department upgrade seminar. We are also benefited from the comments of the participants in the 2021 NASMES, AMES, and IAAE conferences. We thank Jeff Rowley for excellent research assistance. We gratefully acknowledge financial support from ERC grant (number 715940) and the
ESRC Centre for Microdata Methods and Practice (CeMMAP) (grant number RES-589-28-0001).}}

\maketitle
\begin{abstract}
How to allocate vaccines over heterogeneous individuals is one of the important policy decisions in pandemic times. This paper develops a procedure to estimate an individualized vaccine allocation policy under limited supply, exploiting social network data containing individual demographic characteristics and health status. We model the spillover effects of vaccination based on a Heterogeneous-Interacted-SIR network model and estimate an individualized vaccine allocation policy by maximizing an estimated social welfare (public health) criterion incorporating these spillovers. While this optimization problem is generally an \textit{NP-hard} integer optimization problem, we show that the SIR structure leads to a submodular objective function, and provide a computationally attractive greedy algorithm for approximating a solution that has a theoretical performance guarantee. Moreover, we characterise a finite sample welfare regret bound and examine how its uniform convergence rate depends on the complexity and riskiness of the social network. In the simulation, we illustrate the importance of considering spillovers by comparing our method with targeting without network information.

\end{abstract}
\newpage
\section{Introduction}
Allocation of a resource over individuals who interact within a social network is an important task in many fields, such as economics, medicine, education, and engineering (\citet{lee2020key}, \citet{banerjee2013diffusion}, among others). One of the important policy decisions of this sort in pandemic times is how to allocate vaccines over heterogeneous individuals to control the spread of disease and protect the lives of vulnerable. It is crucial for the vaccine allocation rule to take into account the spillover effect of cutting transmission of the disease.

Since the start of COVID-19 pandemic, governments around the world have gone to great lengths to collect network data in which one can trace who is contacting whom. 
Motivated by these observations, we study how to estimate optimal individualized allocations of vaccines under capacity constraint, using micro-level social network data. Data is informative about the covariates of $N$ units, their health status, and their associated neighbors. Using in-sample information, we evaluate the risk to each unit, calculated from its own covariates and spillovers from its heterogeneous neighbors, using an \textit{individualized Susceptible-Infectious-Recovered} model. The purpose of vaccine allocation is to maximize public health, by selecting units to be vaccinated. Obtaining an optimal assignment is, however, challenging since whether a treatment is optimal for an individual depends on which treatments are given to her neighbors. This implies that the search for an optimal allocation has to be performed over the entire network jointly, not individually.
 
This paper makes two main contributions. The first contribution is to develop methods to estimate vaccine assignment policies that exploit network information at the micro-level. The second contribution is to show that the empirical welfare criterion built upon the SIR spillover structure delivers a submodular objective function, which we exploit to obtain computationally attractive algorithms to solve the welfare optimization problem.  Distinct from the existing approach of estimating individualized allocation policies under network interference \citep{viviano2019policy,Abhishek2020OptimalTreatment}, our setting does not assume the availability of Randomized Control Trial (RCT) data. Instead, we assume the availability of estimated values of these spillover parameters from other sources, which we plug into our SIR model. Exploiting already estimated SIR parameter values for immediate targeting and allocation is useful when time is of the essence and the need for policy action is pressing, and avoids the cost of running an RCT.

To optimize the empirical welfare of allocation policies, one naive approach is to evaluate the value of empirical welfare exhaustively for all possible combinations of vaccine allocations over individuals. We refer to this as the brute-force approach. Although the brute-force approach is guaranteed to optimize the empirical welfare, it is not practicable since the number of possible combinations grows exponentially as the number of individuals in the network increases. On the other hand, giving up on optimization entirely and implementing random allocation is indeed practicable, but leads to a significant waste of the vaccine supply, which we show in our simulation exercises. 

Given the challenge in optimizing the empirical welfare, what we recommend in this paper is an allocation policy obtained by greedy optimization. A greedy optimization algorithm in the current setting is to sequentially allocate a vaccine to an individual in the network who is most influential for improving the social welfare. In general, greedy algorithms are not guaranteed to yield an optimum. With the current welfare criterion built upon the SIR spillover structure, however, we can obtain a non-decreasing submodular objective function. Relying on the seminal result in discrete convex analysis shown by \citet{nemhauser1978analysis}, we show that the greedy algorithm delivers an allocation policy at which the value of the objective function is worse than the optimum only up to a universal constant factor, independent of the spillovers, size, and density of the SIR networks. Our derivation of the population welfare regret of the greedily estimated allocation policy reflects the potential loss of welfare due to non-feasibility of obtaining the brute-force allocation policy.

We further illustrate the advantages of our method in our simulation exercises. In a small network setting (up to 35 individuals in the network), comparisons with the brute-force allocation rules reveal that our proposed greedy allocation rules leads to an optimal solution. In a large network setting, we evaluate the performance of our method versus two different assignment rules$\colon$ random assignment, and targeting without considering network information. The welfare improvement relative to these two baselines ranges over 4\% - 12\%, and this result is insensitive to the values of SIR parameters and the size and density of network.

To assess how uncertainty in the SIR parameter estimates affect the welfare performance of the estimated policy, we derive a uniform upper bound of the welfare regret of our vaccine allocation rule and its convergence rate with respect to the size of the sample used for obtaining the SIR parameter estimates. 
The uniform upper bound of regret depends upon two things. Firstly, $n$, which is the sample size of the separate dataset used to estimate the SIR parameters. Secondly, the ratio of the network data sample size $N$ to the maximum number of neighbors $N_M$ plus the minimum between the number of infected units $N_I$ and the number of available vaccine doses $d$ (i.e., $(d\min\{N_M,d\}+2dN_M+\min\{N_I,d\})/N$). As $N_M$ and $N_I$ grow, the complexity and risk of the social network increase, which can reduce the welfare regret performance of the estimated vaccine allocation rule.

The remainder of this paper is organized as follows. We first discuss the relevant literature in the rest of this section. Section 2 details various models, and the HI-SIR model in particular, and the wider setting. Section 3 is concerned with estimation, including the estimation of SIR parameters and the construction of the QIP problem. The optimization procedure is contained in section 4. Section 5 contains the theoretical results. Simulation details are shown in Section 6, and Section 7 concludes. All proofs and derivations are shown in the appendix.

\subsection{Related Literature}
Our work contributes to the literature on statistical treatment rules, which was first introduced into econometrics by \citet{manski2004statistical}. The optimal treatment allocation regime has been studied in many fields, such as medical statistics \citep{zhao2012estimating,zhao2015new}, operational research \citep{loiola2007survey} and economics. Following the pioneering works of \citet{hannan1957approximation} and \citet{savage1951theory},\footnote{\citet{hannan1957approximation} considers regret-minimizing strategies in the context of zero-sum and sequential games. \citet{savage1951theory} introduces minimax-regret rules to the statistical decision theory.} researchers in econometrics and machine learning often use regret to evaluate the performance of decision rules. The recent literature of statistical treatment rules includes \citet{dehejia2005program}, \citet{hirano2009asymptotics}, \citet{stoye2009minimax,stoye2012minimax}, \citet{tetenov2012statistical}, \citet{bhattacharya2012inferring}, \citet{kitagawa2018should}, \citet{zhou2018offline}, \citet{manski2019treatment}, \citet{kasy2019adaptive}, \citet{athey2020policy}, \citet{kock2020functional}, \citet{mbakop2021model}, \citet{manski2021statistical}, \citet{sakaguchi2021estimation}, and \citet{kitagawa2021constrained} among others. The planner's objective function in the majority of these works is a sum of individual outcomes under the no-interference assumption (i.e., Stable Unit Treatment Value Assumption of \citet{rubin1974estimating}). This assumption does not hold in this study because of the network spillover effects that are present. To our knowledge, there are only two other papers that also consider the network setting in statistical treatment choice, which are \citet{viviano2019policy} and \citet{Abhishek2020OptimalTreatment}. These two papers assume the availability of pilot data from RCT studies performed over networks in order to form empirical welfare criteria. Their frameworks are not restricted to the SIR setting of the current paper and cover spillover structures commonly assumed in social science applications. In contrast, our approach forms welfare estimates by imposing the HI-SIR model structure and plugging in values of the primitive spillover parameters that are estimated or calibrated in some external study (e.g., \citet{baqaee2020reopening}). Another notable difference is that we consider allocation policies that are not constrained other than via the capacity constraint, while \citet{viviano2019policy} and \citet{Abhishek2020OptimalTreatment} assume the class of implementable allocation policies has a finite VC-dimension to control overfitting to the training RCT sample.
 
The SIR model was originally proposed by \citet{kermack1927contribution}, and is now the workhorse model in the epidemiological literature. Many extended versions have been studied in epidemiological analyses, such as the Susceptible-Infected-Susceptible model \citep{naasell1996quasi} and the Susceptible-Exposed-Infected-Recovered model \citep{li1995global}. During the global pandemic, an epidemological literature has sprung up within economics. \citet{atkeson2020will} and \citet{stock2020data} introduced the SIR model into economics to study the implications of the current pandemic on the US economy. We introduce heterogeneity into the SIR model, which is similar to what  \citet{acemoglu2020multi} does in studying the Multi-Risk SIR model. That paper assumes, however, that the infection rate after the release of a vaccine equals zero, which means it does not consider the vaccine allocation problem. Our work contributes to the current literature by studying micro-level vaccine assignment rules in a heterogeneous SIR model with network information. In contrast, the existing works analyzing vaccine allocation rules focus on solving for the optimal proportion of vaccinated units in the population (\citet{pastor2002immunization}, \citet{manski2010vaccination,manski2017mandating}). \citet{chen2020allocation} analyzes vaccine allocation using a heterogeneous SIR model, while they consider vaccine allocation policies at the group-level rather than the individual-level.
 
We build a connection to the literature on using a submodular function to solve an optimization problem. The performance guarantee of a general greedy algorithm for solving submodular maximization problems with a cardinality constraint was first established by \citet{nemhauser1978analysis}. The later literature links the cardinality constraint to a more general constraint$\colon$ Matroid constraint \citep{fisher1978analysis,cunningham1985minimum}. See \citet{bach2011learning} and \citet{krause2014submodular} for overviews of papers studying optimization of submodular functions. In this work, we discuss a submodular function with a uniform matroid constraint (i.e., capacity constraint) and a more general partition matroid constraint.

We notice that our approach to vaccine allocation problem is related to the influence maximization problem first formulated by \citet{kempe2003maximizing}. \citet{chen2010scalable} investigates submodularity of objective functions and greedy optimization algorithms in this problem. Applications of the influence maximization problem include targeting for viral marketing (\citep{domingos2001mining}) and optimal information spread in social network (\citep{bakshy2011everyone}). There are two widely studied information diffusion models in this literature: \textit{Independent Cascade Model} \citep{goldenberg2001talk} and \textit{Linear Threshold Model} \citep{granovetter1978threshold}. Despite some similarity between the diffusion models and our SIR model, this literature has not considered individualized vaccine allocation problem.

We also note that there is a growing literature on estimation of treatment effects under network interference. \citet{manski2013identification} discusses identification of treatment effects and spillover effects under a deterministic interference graph and a set of relevant potential outcomes. The increased number of network datasets that have recently become available has motivated further work on this topic, including \citet{savje2017average}, \citet{aronow2017estimating}, \citet{athey2018exact}, \citet{basse2019randomization}, and \citet{leung2019normal}. \citet{li2020random} non-parametrically estimates direct and indirect effects of treatment in a random network setting. \citet{vazquez2020causal} analyzes estimation of spillover effects using an instrument variable. See \citet{kline2020econometric} and \citet{graham2020econometric} for recent reviews on econometric analysis in the presence of social interactions. 

\section{Setup and Identification}
We consider a basic model to study the vaccination allocation problem. Let us first introduce the timeline and data setting that we consider in this work. 
\begin{center}
\begin{tikzpicture}[
    every node/.style = {align=center},
          Line/.style = {-angle 90, shorten >=2pt},
    Brace/.style args = {#1}{semithick, decorate, decoration={brace,#1,raise=2pt,
                             pre=moveto,pre length=2pt,post=moveto,post length=2pt}},
            ys/.style = {yshift=#1}
                    ]
\linespread{0.8}                        
\coordinate (a) at (0,0);
\coordinate[right=10mm of a]    (b);
\coordinate[right=20mm of b]    (c);
\coordinate[right=50mm of b]    (d);
\coordinate[right=50mm of d]    (e);
\coordinate[right=10mm of e]    (f);
\draw[Line] (a) -- (b) -- node[above] {$1^{st}$ period}(d)-- node[above] {$2^{nd}$ period} (e) --  (f) node[below right] {time};
\draw[Line] (b) ([ys=-7mm]  b) node[below] {t=0} -- (b);
\draw[Line] (d) ([ys=-7mm]  d) node[below] {t=1} -- (d);
\draw[Line] (d) ([ys=10mm]  d) node[above] {Vaccine} -- (d);
\draw[Line] (e) ([ys=-7mm]  e) node[below] {t=2} -- (e);
\draw[Brace=mirror] (b) -- node[below=6pt] {$A$}(c);
\draw[Brace=mirror] (c) -- node[below=6pt] {$[X_i, H_{0i}]_{i=1}^N$}(d);
\draw[Brace=mirror] (d) -- node[below=6pt] {$[H_{1i}]_{i=1}^N$}(e);
\end{tikzpicture}
\end{center}
As shown in the illustration, we suppose there are two periods. At $t=0$, policymakers initially observe the network structure $A$ (i.e., adjacency matrix) linking $N$ individuals, for which we provide further details below. Policymakers then observe covariates $X_i\in\mathcal{X}\subset\mathbbm{R}^{d_x}$ and current period health state $H_{0}\in\{S,I,R\}$ for each of the $N$ individuals. The health states $\{S,I,R\}$ stand for \textit{Susceptible, Infected}, and \textit{Recovered}. We assume the network structure A is observed before personal health status to avoid the impact of self-isolation on the network structure. At $t=1$, policymakers start to assign the vaccine. After a short vaccination period, people begin to meet their neighbors, which we call the interaction period. The health state during that period is defined as $H_1\in\{S,I,R,D\}$, where $D$ stands for death. Since at the time of assigning vaccination, $H_1$ is not yet observed by researchers, a stochastic health state will be used to evaluate personal risk. The ultimate goal of policymakers is to maximize the expected social health situation via the allocation of vaccines.

In our setting, units are connected through a social network. We assume the following property on network structure holds$\colon$
\begin{assumption}{(Undirected Relationships)}\label{assumpundir}
The interference graph is undirected. i.e., $A_{ij}=A_{ji}.$
\end{assumption}
\noindent The symmetric $N\times N$ adjacency matrix A specifies who contacts with whom, with the $(i,j)$th element of A, denoted by $A_{ij}$, equal to one if unit $i$ and unit $j$ has positive contact time, and zero otherwise. By convention, all the diagonal elements $A_{ii}$ are equal to zero. If $A_{ij}=1,$ then we say that $i$ and $j$ are neighbors. Let $N_i$ indicate the neighbors of unit $i$, then we write $A_{ij}=1$ if $j\in N_i$ and $i\in N_j$.
The size of spillover (i.e., the probability of disease transmission) between the units $i$ and $j$ depends not only on $A_{ij}$ but also on the amount of their contact time and the transmission rates which are allowed to be asymmetric between them. We accordingly have a directed weighted network structure for the spillovers, as shown in later sections.

Now, let us introduce the notation that we use in the following sections. First, $v_i$ is the individual vaccine assignment rule (i.e., $v_i=1$ if unit $i$ gets the vaccine). Let $\mathbf{v}$ denote $(v_1,...,v_N)\in \{0,1 \}^{N},$ and $X$ denote $(X_1,...,X_N)\in \mathbbm{R}^{N\times d_x}.$ Let $S_i$ be the susceptible state indicator in the first period (i.e., $S_i = \mathbbm{1}_{\{H_{0i}= S\}}$), let $I_i$ be the infected state indicator in the first period (i.e., $I_i = \mathbbm{1}_{\{H_{0i}= I\}}$), and let $R_i$ be the recovered state indicator in the first period (i.e., $ R_i = \mathbbm{1}_{\{H_{0i}= R\}}$). Moreover, let $\abs{N_i}$ denote the number of neighbors of unit $i$ (i.e., $\abs{N_i} = \sum_j A_{ij}$).

\subsection{Heterogeneous-Interacted-SIR model}\label{section 2.2}
To measure the personalized transition probability, we use a \textbf{HI-SIR} model. Our model is defined in discrete time within a simplified setting of two time periods. In the first period, we observe the health state of each unit $H_{0},$ which belongs to \textit{S}(Susceptible), \textit{I}(Infected), or \textit{R}(Recovered),
\begin{equation}
    S_i+I_i+R_i = 1.
\end{equation}
In the second period, the state variable is $H_{1}$. Compared with $H_{0},$ $H_{1}$ includes one more state \textit{D}(Death). 
\begin{equation}
    \mathbbm{1}_{\{H_{1i}= S\}}+\mathbbm{1}_{\{H_{1i}= I\}}+\mathbbm{1}_{\{H_{1i}= R\}}+\mathbbm{1}_{\{H_{1i}= D\}} = 1.
\end{equation}
Without the vaccine, the state can move from susceptible to infected, then to either recovery or death. Now, we consider the setting after introducing the vaccine. Generally, vaccination has two purposes$\colon$ the first is limiting the spread of disease, and the second is treatment. Vaccination builds up the immune system, which leads to recovery. However, the effectiveness of vaccination (i.e., the percentage of vaccinated units that recover) is not clear. For simplicity, we assume that assumption \ref{assumpperfect} holds.
\begin{assumption}\label{assumpperfect}
(PT) Perfect Treatment$\colon$
A vaccinated unit enters the Recovered state, regardless of its previous state (i.e., $\Pr(H_{1i}=R\vert v_i=1)=1$).
\end{assumption}

To further simplify the setting, we split all units into a finite number of disjoint groups based on their characteristics. The infection rate between each group varies. This setting could be extended to the individual level, but the micro level infection rate would need to be known in this case. Here, we consider two groups and use \textbf{age} as a binary indicator$\colon$ $G_1$ (\textit{Young}) and $G_2$ (\textit{Old}). We now define $a_i$ and $b_i$ as the group indicators (i.e., $a_i = \mathbbm{1}_{\{X_i\in G_1\}}$ and $b_i = \mathbbm{1}_{\{X_i\in G_2\}}$).

We specify one of the key components in SIR models, the infection rate of unit $i$, as$\colon$
\begin{equation}\label{eq_infect}
\begin{split}
    q_i &= \Big[\beta_{11}\frac{\sum_{j\in N_i}I_j (1-v_j)a_j}{|N_i|} +\beta_{12}\frac{\sum_{j\in N_i} I_j (1-v_j) b_j }{\abs{N_i}}\Big]\cdot a_i\\
   & +\Big[\beta_{21}\frac{\sum_{j\in N_i} I_j(1-v_j) a_j }{\abs{N_i}}+\beta_{22}\frac{\sum_{j\in N_i}I_j (1-v_j)b_j}{\abs{N_i}}\Big]\cdot b_i,\\
    \end{split}
\end{equation}

\noindent where $\beta_{sk}=-\kappa_s\ln(1-c_{sk}),$ $c_{sk}$ is the probability of successful disease transmission following a contact between group $s$ and group $k$ (i.e., $c_{11}$ measures the transmission probability from one unit to another within the young group, $c_{12}$ is the corresponding probability of transmission from a unit in the old group to a unit in the young group, with similar definitions for $c_{21}$ and $c_{22}$), and $\kappa_s$ is the average number of contacts in group $s$ at each time period. $\beta_{sk}$ describes the effective contact rate of the disease between group $s$ and group $k$. The derivation of equation \ref{eq_infect} can be found in the appendix.

In the above expression, $I_j(1-v_j)$ means a susceptible individual can only be infected by neighbors who were infected and not vaccinated.\footnote{This can also be thought of as an underlying assumption, which is commonly used in the epidemiological literature (e.g., \citet{hashem2020match}).} Those neighbors may come from various groups. We calculate the fraction of neighbors in each group and multiply them by the associated risk parameters. The risk parameter $\beta_{sk}$ measures the probability that a susceptible individual in group $k$ is infected by an infected individual from group $s$ in one time period. 

We now define $\{\gamma_1, \gamma_2\}$ as the recovery rate and $\{\delta_1,\delta_2\}$ as the mortality from infection in group 1 and group 2 respectively. Given this, we can formulate the probability of staying in the infection state for the infected unit $i$ as$\colon$
\begin{equation}\label{eq:recov}
p_i = 1-a_i(\gamma_1+\delta_1)-b_i(\gamma_2+\delta_2)  .  
\end{equation}
Since the probability of recovery and death purely depend on personal physical fitness,\footnote{This can be thought as a simplified assumption, which indicates the death rate does not depend on the availability of hospital spare capacity.} there is no interactive part in equation \ref{eq:recov}. The transition probability to the infected state is then$\colon$
\begin{equation}\label{eq:8}
   \begin{split}
P_{Ii} (\mathbf{v}) \equiv \Pr(H_{1i} = I\vert  X, \mathbf{v}, A, H_{0})= S_iq_i\cdot(1-v_i) + I_ip_i\cdot (1-v_i).
   \end{split}
   \end{equation}
In the above expression, the probability of an unvaccinated unit being infected has two components. The first is the probability of a healthy unit being infected. The second is the probability of staying in the infected state for those infected in the first period. Under Assumption \ref{assumpperfect}, a vaccinated unit has zero probability of being infected. Similarly, the transition probability to the susceptible state is$\colon$
\begin{equation}
    P_{Si} (\mathbf{v}) \equiv \Pr(H_{1i} = S\vert  X, \mathbf{v}, A, H_{0})= \big[1-v_i-q_i(1-v_i)\big]\cdot S_i.
\end{equation}
An unvaccinated unit can only exit the susceptible state by infection. Therefore, the probability of staying in the susceptible state decreases with the risk parameter $\beta_{sk},$ which depends on the number of infected neighbors and the number of contacts with them. The remaining two states do not rely on the network structure. First, the transition probability to the recovered state $\colon$
\begin{equation}
\begin{split}
    P_{Ri}(\mathbf{v}) \equiv \Pr(H_{1i} = R \vert  X, \mathbf{v}, A, H_{0})= 
    v_i+\big[R_i+I_i(a_i\gamma_1+b_i\gamma_2)\big]\cdot(1-v_i).
    \end{split}
\end{equation}
In the above expression,\footnote{A maintained assumption in this equation is that the probability of being reinfected for the recovered units is zero. We relax this assumption in Section 7.} recovery has two different sources. One is the vaccine, and the other is self-immunity. The effect of self-immunity is heterogeneous and varies with personal characteristics. The probability of building immunity in each group is $\gamma_1,\gamma_2.$ The last state is death, which occurs with probability
\begin{equation}
    P_{Di}(\mathbf{v}) \equiv \Pr(H_{1i} = D \vert  X, \mathbf{v}, A, H_{0}) =I_i(a_i\delta_1+b_i\delta_2)\cdot(1-v_i).
\end{equation}
\subsection{Optimal Vaccine Allocation Problem}
In \citet{emanuel2020ethical}, a group of medical ethics experts suggest a successful vaccine is needed to reduce death and morbidity from infection, and is also needed for the restoration of economic and social activity. Following that suggestion, we choose our baseline outcome variable as the weighted average of the probability of being healthy in the second period. The idea of using weighted probability is to allow a flexible policy target of the planner. For example, if the planner wants to incorporate the importance of economic recovery into the policy objective, she may want to weight more the probabilities of being healthy of those who can contribute more to the economic output. For instance, the planner could specify the weights on the individuals to depend on their individuals characteristics including working hours and other socioeconomic characteristics (i.e., $g_i=g(X_i)$).\footnote{A maintained assumption in this expression is that for every unit, the weight is same for both susceptible and recovered states. This is a simplifying assumption that can be relaxed if we want to weight differently the susceptible and recovered states.} We assume the weight is non-negative for every unit. Taking these into consideration, equation (\ref{eq:1}) specifies the goal of the vaccine allocation policy as a constrained optimization problem:
\begin{equation}\label{eq:1}
           \max_{\mathbf{v}\in \{0,1 \}^N}\frac{1}{N}\sum_{i=1}^{N}g_i\sum_{h\in \{S,R\} } P_{hi}(\mathbf{v}),
\end{equation}
s.t.
\begin{equation}
    \sum_{i=1}^{N} v_i \leq d,
\end{equation}
where
\begin{equation}
    P_{hi}(\mathbf{v}) = \Pr(H_{1i}=h\vert X,\mathbf{v}, A, H_0),
\end{equation}
and $d \geq 1$ is a positive integer for the exogenous cardinality constraint. The main idea of the above objective function is to maximize the weighted probability of being in the susceptible or recovered state in the second period by appropriately assigning the $d$ doses of vaccine at the end of the first period.

In equation (\ref{eq:1}), $P_{hi}$ is the heterogeneous state transition function, which describes the probability of $h\in\{S,R\}$ in the second period. This transition probability depends on the individuals' covariates and previous state including whether being vaccinated or not, and the associated network structure. We adopt the HI-SIR model to formulate the above transition function, which has been provided in the previous subsection.

One relevant question is$\colon$ \textit{Will vaccine allocation change the network structure}? Yes, it would change the behaviour of vaccinated units. For example, vaccinated units prefer to go out as compared to unvaccinated units. Given this, the number of contacts at each time period $\kappa_s$ and the network structure $A$ would change after the vaccine allocation. Our framework allows the network structure to vary without affecting the optimal allocation of vaccines in a special case where \textit{only the vaccinated units} change their behaviours. This is because under our perfect treatment assumption, the vaccinated units no longer spread the disease or be infected, and their behavioral changes do not affect the health statuses of the neighbors and themselves. On the other hand, our framework cannot accommodate a general case where the unvaccinated units also change their behaviours, since if so the heterogeneous SIR parameters in the objective function change in response to the vaccine allocation. To allow this scenario, we could incorporate uncertainty as to the values of $\kappa_s$ and $A$ in the second period, for instance, by optimizing an objective function that takes the expectation of the SIR parameters the adjacency matrices conditional on $\mathbf{v}$. We do not, however, consider such an extension in this paper and leave this topic for future research.

\section{Estimation}
In order to measure the individual risk level using the HI-SIR model, we need to know the associated SIR parameters$\colon$ transmission rate (i.e., $\beta_{11},\beta_{12},\beta_{21},\beta_{22}$), and recovery rate (i.e., $\gamma_1,\gamma_2$). Given that we cannot observe the true value of those parameters, it is infeasible to evaluate the objective function (\ref{eq:1}) based on the in-sample information of $(H_0,X)$ and $A$ of the target network. We therefore assume access to a separate dataset with sample size $n$ or an external study analyzing it, from which we can form estimates for these exogenous parameters. We construct an empirical version of the population welfare (\ref{eq:1}) and maximize over the feasible allocation policies. To reflect the precision of the SIR parameter estimates in the welfare performance of an estimated allocation rule, we explicitly take into account the sampling uncertainty of the parameter estimates in our derivation of the welfare regret upper bound.

\subsection{Estimation of SIR Parameters}
The estimation of infection rate and death rate always faces severe missing data problems as discussed in \citet{manski2020estimating}. \citet{keeling2011modeling} points out that, usually, researchers first estimate the reproductive ratio $\mathcal{R}_0$, which is the average number of individuals that one sick person infects.
\begin{equation}
    \mathcal{R}_0 = \beta\times \frac{1}{\gamma}.
\end{equation}
Then, the infection rate can be derived from the estimated recovery rate $\hat{\gamma}$ and $\hat{\mathcal{R}}_0$. In our case, the reproductive ratio is heterogeneous at group level.  
\begin{equation}\label{eq:r0}
    \mathcal{R}_{0sk} = \beta_{sk}\times \frac{1}{\gamma_{s}} \quad \forall s,k = 1,2, 
\end{equation}
where $\mathcal{R}_{0sk}$ is the number of infectious individuals in group $s$ resulting from one sick person in group $k$. We need to estimate the average number of younger infectious and older infectious from one sick person in group 1 and group 2, and also the recovery rate in each group. Given these values, we can estimate $\beta_{11},\beta_{12},\beta_{21},\beta_{22}$ from equation (\ref{eq:r0}).

\begin{remark}
We do not discuss what is a desirable procedure for estimating the model parameters in this work, since the choice of estimator depends on the type of data (e.g., Seroprevalence data, Reported cases data, etc.). See \citet{keeling2011modeling} for further details. For the COVID-19 transmissions, estimation of homogeneous $\mathcal{R}_0$ and other SIR parameters has been performed in several papers including \citet{fernandez2020estimating}, \citet{ferguson2020report}, and \citet{korolev2021identification}. They note the difficulty in calibrating critical parameters at an early stage of the pandemic due to the lack of credible data, which motivates partial identification analysis of \citet{manski2020estimating} and \citet{stoye2021bounding}. Our approach, however, assumes availability of credible point estimates and does not allow identified-set estimates for the SIR parameters. See \citet{ellison2020implications} and \citet{akbarpour2020socioeconomic} for recent estimates of heterogeneous SIR parameters. 
\end{remark}

\subsection{Quadratic Integer Programming}

Plugging the parameter estimates into our \textbf{HI-SIR} model, we now have the sample analog of the population maximization problem (\ref{eq:1}), which is
\begin{equation}\label{eq:17}
\begin{split}
&\max_{\mathbf{v}\in \{0,1 \}^N}\mathcal{W}_n(\mathbf{v}), \quad  \text{s.t.} \quad \sum_{i=1}^N v_i \leq d, \hspace{4em} \text{where}\\
&\mathcal{W}_n(\mathbf{v}) =\frac{1}{N}\sum_{i=1}^{N}g_i\sum_{h\in\{S,R\}} \hat{P}_{hi}(\mathbf{v}).
\end{split}
\end{equation}

We can formulate this optimization as a quadratic integer programming (\textbf{QIP}) problem, which in the context of an assignment problem over a network is synonymous with the Quadratic Assignment Problem (\textbf{QAP}) of \citet{koopmans1957assignment}. We can express $\mathcal{W}_n(\mathbf{v})$ as 
\begin{spacing}{0.9}
\begin{equation}\label{eq:QIP}
\begin{split}
    \mathcal{W}_n(\mathbf{v})&=\frac{1}{N}\sum_{i=1}^{N} g_i\underbrace{\bigg[v_i +\Big[R_i+(a_i\hat{\gamma}_1+b_i\hat{\gamma}_2) I_i\Big](1-v_i)
    +M_iS_i(1-v_i)\bigg]}_{\textit{Probability of being healthy}}
\end{split}
\end{equation}
where
\begin{equation}
  M_{i} =1- \frac{\sum_{j=1}^{N}(\hat{\beta}_{11}a_ia_j+\hat{\beta}_{12}a_ib_j+\hat{\beta}_{21}b_ia_j+\hat{\beta}_{22}b_ib_j)A_{ij}I_j(1-v_j)}{\abs{N_i}}.
\end{equation}
\end{spacing}
\vskip 0.3cm
For the probability of being healthy in equation (\ref{eq:QIP}), there are two linear terms and one quadratic term in $\mathbf{v}$. The first term measures the direct effect of vaccination. A vaccinated unit is safe from infection with $100\%$ probability. The last two terms describe the probability of being free of infection for unvaccinated units. Infected units naturally recover with probability $\{\gamma_1, \gamma_2\}$, which depends on their own characteristics. For those units who are already recovered in the first period, they are free from infection in the second period. The last component takes into account the indirect effect of vaccination. For susceptible units, the probability of being infected by their infected neighbors is summarized by the interaction term.

After removing all the constant parts in equation (\ref{eq:QIP}), we obtain a simplified objective function (i.e., $\mathcal{W}_n(\mathbf{v})= F_n(\mathbf{v}) + constant$)$\colon$
\begin{equation}\label{eq:23}
\begin{split}
    F_n(\mathbf{v}) = \sum_{i=1}^N\hat{c}_{i}v_{i}+\frac{1}{N}\sum_{i=1}^{N}\frac{1}{\abs{N_i}}\sum_{j=1}^{N}T_{ij}(v_i+v_j-v_iv_j),
    \end{split}
\end{equation}
where
\begin{equation}
    \hat{c}_i = g_i\big[1-R_i-(a_i\hat{\gamma}_1+b_i\hat{\gamma}_2)I_i-S_i\big]/N,
\end{equation}
\begin{equation}
    T_{ij} = g_i\big(\hat{\beta}_{11}a_ia_j+\hat{\beta}_{12}a_ib_j+\hat{\beta}_{21}b_ia_j+\hat{\beta}_{22}b_ib_j\big)A_{ij}I_jS_i.
\end{equation}
\vskip 0.3cm
Since $F_n$ differs from $\mathcal{W}_n$ only by an additive constant (conditional on the network structure and individual characteristics in the first period), maximizing $F_n$ is equivalent to maximizing the original empirical welfare function $\mathcal{W}_n$. Therefore, from now on, we will focus on $F_n(\mathbf{v})$ as our new objective function. Within $F_n(\mathbf{v})$, there is a quadratic term plus linear components in $\mathbf{v}$. Current software is available to solve general QIP problems, such as CPLEX and Gurobi. However, both applications require a \textbf{symmetric weighting matrix}, which does not hold in our case. This asymmetric property comes from the infectious process, since disease can only be transmitted from infected units to susceptible units, but the reverse is not true. We discuss how to solve this QIP problem with showing and exploiting the submodular property of our objective function in the next section.


\section{Optimization}
\subsection{Submodularity}
We showed in the last section that we can formulate our objective function as QAP. This kind of problem is well known as an \textit{NP-hard} and \textit{NP-hard to approximate} problem \citep{cela2013quadratic}. In general, we cannot solve QAP in polynomial time, which is an issue in practice. We shall, however, show that the quadratic integer programming in our vaccine allocation problem can be linked to the submodular optimization problem. The benefit of submodularity is that there exist off-the-shelf algorithms that can solve a submodular minimization problem in exact polynomial time and \textit{approximately} solve a submodular maximization problem with capacity constraint in polynomial time. The seminal result of \citet{nemhauser1978analysis} provides a universal bound for the quality of approximation as detailed below in Section \ref{section4.2}.

\begin{definition}[Submodular function]
Let $\mathcal{N}=\{1,2, \dots, N \}$. A real-valued set-function $F\colon 2^\mathcal{N}\rightarrow \mathbb{R}$ is submodular if and only if, for all subsets $A,B	\subseteq \mathcal{N},$ we have$\colon$ $ F(A)+F(B)\geq F(A\cap B)+F(A\cup B).$
\end{definition}

In simple terms, submodularity describes the diminishing returns property. The marginal increase in the average probability of being healthy decreases in the number of vaccinated units. This property is crucial for the maximization algorithm. For ease of exposition, we express the simplified empirical welfare $F_n$ as a set function with argument $V \in 2^{\mathcal{N}}$, where the binary vector of vaccine allocation $\mathbf{v} \in \{0,1 \}^N$ and $V$ correspond by $V = \{i \in \mathcal{N}: v_i = 1 \}$:
\begin{equation}\label{eq:26}
F_n(V) = \mathbf{v} ^\intercal \hat{W} \mathbf{v} + \hat{C}^\intercal \mathbf{v} - \mathbf{1}_{N\times 1}^\intercal \hat{W} \mathbf{v} - \mathbf{v} ^\intercal \hat{W} \mathbf{1}_{N\times 1},
\end{equation}
where
\begin{equation}
\footnotesize
    \hat{C}=
  \begin{bmatrix}
\hat{c}_1\\
    \vdots\\
    \hat{c}_N
    \end{bmatrix},
\hspace{4em}
    \mathbf{1}_{N\times 1}=
    \begin{bmatrix}
    1\\
    \vdots\\
    1
    \end{bmatrix},
\hspace{4em}
\footnotesize
    \hat{W} = 
    \begin{bmatrix}
    \hat{w}_{11} & \hat{w}_{12} & \cdots\  & \hat{w}_{1N}  \\
    \hat{w}_{21} & \hat{w}_{22} & \cdots\  & \hat{w}_{2N}  \\
    \vdots & \vdots & \ddots   & \vdots  \\
    \hat{w}_{N1} & \hat{w}_{N2} & \cdots\  & \hat{w}_{NN}  \\
    \end{bmatrix},
\end{equation}
\begin{equation}
    \hat{w}_{ij} = -\frac{A_{ij}g_i}{\abs{N_i}N}(\hat{\beta}_{11}a_iS_ia_jI_j+ \hat{\beta}_{12}a_iS_ib_jI_j +\hat{\beta}_{21} b_iS_ia_jI_j +\hat{\beta}_{22}b_iS_ib_jI_j).
\end{equation}
We then denote the class of feasible allocation sets $V$ subject to the cardinality constraint $|V| \leq d$ by $\mathcal{V}_d \equiv \{V \in 2^{\mathcal{N}} : |V| \leq d \}$. Since vaccinating additional individuals cannot reduce welfare, $F_n$ is a \textit{non-decreasing} set function, i.e., for any $V \subset V'$, $F_n(V) \leq F_n(V')$.\footnote{See the proof of Theorem \ref{corollary4.1} for a formal proof for the non-decreasing property of $F_n$.} 

The quadratic functional form of $F_n$ shown in (\ref{eq:26}) can be linked to one classic submodular function called a \textit{cut function}. Cut functions have been well studied in combinatorial optimization and graph theory. We apply some of the results from that literature (e.g., \citet{bach2011learning}).

\begin{lemma}\label{proposition3.1}
Let $\hat{W}\in\mathbb{R}^{N\times N}$ and $\hat{C}\in\mathbb{R}^{N}.$ Then the set function $F_n\colon V \mapsto  \mathbf{v} ^\intercal \hat{W} \mathbf{v} + \hat{C}^\intercal \mathbf{v} - \mathbf{1}_{N\times 1}^\intercal \hat{W} \mathbf{v} - \mathbf{v} ^\intercal \hat{W} \mathbf{1}_{N\times 1}$ is submodular if and only if $\hat{w}_{ij}\leq 0\ \forall i\neq j.$
\end{lemma}

The proof is shown in the appendix. Note that the necessary and sufficient condition for submodularity shown in this lemma is distinct from negative semidefiniteness of the matrix $\hat{W}$. Since all the parameters in $\hat{w}_{ij}$ are non-negative, we must have $\hat{w}_{ij}\leq 0, \forall i,j=1,...,N$. This immediately leads to the following theorem:

\begin{theorem}\label{corollary4.1}
The objective function $F_n(V)$ is a non-decreasing submodular function for any adjacency matrix, covariate values, and parameter estimates.
\end{theorem}
Theorem \ref{corollary4.1} is the key result in our paper. It describes two important properties of our objective function; monotonicity and submodularity. We exploit these two properties to justify the uses of greedy maximization algorithms shown in the next subsection.

\subsection{Greedy Maximization Algorithm} \label{section4.2}
Greedy maximization algorithms for submodular functions have been studied and frequently used for well over forty years. The performance guarantee of the algorithm that we study was first introduced by \citet{nemhauser1978analysis}. This algorithm essentially uses the diminishing returns property of the submodular function. The idea is to iteratively select the most valuable element until the capacity constraint is reached. At each round, the algorithm evaluates $\mathcal{O}(N)$ functions to identify the marginal gain of each element. The number of rounds depends on the capacity constraint $d.$ As a result, the computational complexity of the greedy algorithm is of order $\mathcal{O}(N\cdot d)$, well below the computational complexity of the brute-force search. Algorithm \ref{algorithm1} presents the greedy maximization algorithm applied to maximization of the empirical welfare (\ref{eq:26}). 
\vskip 0.4cm
\begin{spacing}{0.8}
\begin{algorithm}[H] \label{algorithm1}
\begingroup
\linespread{1}
\SetAlgoLined
1$\colon$ \textbf{Input$\colon$} Dataset $\{S_{i},I_{i},R_{i},a_{i},b_{i}\}_{i=1}^{N},$ $\{A_{ij}\}_{i,j=1}^{N}$, estimated parameters $\{\hat{\beta}_{11},\hat{\beta}_{12},\hat{\beta}_{21},\hat{\beta}_{22},\hat{\gamma}_{1},\hat{\gamma}_{2}\}$, weight $\{g_i\}_{i=1}^N$ and capacity constraint $d$\;
\vskip 0.2 cm
2$\colon$ \textbf{Initialization}$\colon$ Starting from the empty set $V = \emptyset$ \;
\vskip 0.2 cm
\eIf {$\vert V\vert< d$}{
\vskip 0.2 cm
3$\colon$ \textbf{for} each $i \in \mathcal{N}\backslash V$ \textbf{do}\
\vskip 0.2 cm
4$\colon$ Compute the marginal gain $F_n(V+\{i\})\ -\ F_n(V)$\;
\vskip 0.2 cm
5$\colon$ Select $i$ which maximizes the marginal gain and add it into the set $V$\;
\vskip 0.2 cm
}{
\vskip 0.2 cm
\textbf{return} the set $V$\;
}
\caption{Capacity Constrained Greedy Algorithm}
\endgroup
\end{algorithm}
\end{spacing}

\vskip 0.4cm

In general, there is no performance guarantee of the greedy algorithm. However, as shown by \citet{nemhauser1978analysis} for a non-decreasing submodular function with cardinality constraint (i.e., capacity constraint in our case), the greedy maximization algorithm is guaranteed to yield an allocation rule $\hat{V} \in \mathcal{V}_d$ that satisfies $F_n(\hat{V}) \geq (1-\alpha_d) F_n(\hat{V}^{\ast})$, where $\hat{V}^{\ast} \in \mathcal{V}_d$ is a constrained optimum under the capacity constraint, and $\alpha_d$ is a positive constant that depends only on $d \geq 1$ and $\alpha_d \geq 1/e$ for all $d \geq 1$. This seminal result implies that the greedy maximization algorithm provides a universal optimization guarantee for non-decreasing submodular functions, $F_n(\hat{V}) \geq (1-1/e) F_n(\hat{V}^{\ast}) \approx 0.63 F_n(\hat{V}^{\ast})$. Since we show in Theorem \ref{corollary4.1} that our objective function is non-decreasing and submodular, we obtain the following theorem as an immediate corollary of our Theorem \ref{proposition3.1} and \citet{nemhauser1978analysis}.

\begin{theorem}[\citealt{nemhauser1978analysis}]\label{theorem4.1}
Let $F_n : 2^{\mathcal{N}} \to \mathbb{R}$ be the simplified empirical welfare function as defined in (\ref{eq:26}) and $\hat{V}^* \in \arg \max_{V \in \mathcal{V}_d}F_n(V)$, $d \geq 1$. The greedy algorithm shown in Algorithm \ref{algorithm1} outputs $\hat{V} \in \mathcal{V}_d$ such that
\begin{equation}
    F_n(\hat{V})\geq (1-\alpha_d) F_n(\hat{V}^*) \geq  (1-1/e)F_n(\hat{V}^*),
\end{equation}
where $1-\alpha_d \equiv 1-\left(1 - \frac{1}{d} \right)^d $ is monotonically decreasing in $d$ and converges to $1-e^{-1}$ as $d \to \infty$.
\end{theorem}

\subsection{Targeting Constraint}
Up until now, we have only considered a simple capacity constraint in the vaccine assignment rule. In reality, Beyond the weight specification in the objective function, policymakers may want to prioritize some group over the others by limiting the number of vaccines that are administered in each group.\footnote{A group in this section does not need to coincide with the group defining the heterogeneity of the SIR parameters. For example, we could divide units based on their job category, geographical location, or community.} For example, policymakers may limit access to vaccines for those people that can work at home. If we are able to divide individuals into two groups based on their job categories, into a group that can work at home and a group that cannot say, then policymakers can set an upper bound on the number of vaccines that are available for the work at home group.


We call this kind of constraint a \textit{targeting constraint}, and impose it in our model in such way that each of the two age groups has a capacity constraint for the number of available vaccines:
 \begin{equation}
     \sum_{i\colon X_i\in G_1}v_i=\sum_{i=1}^{N}a_iv_i\leq d_1,
     \hspace{4em}
     \sum_{i\colon X_i\in G_2}v_i=\sum_{i=1}^{N}b_iv_i\leq d_2.
 \end{equation}
This targeting constraint belongs to a general class of constraints$\colon$ the so called \textit{matroid} class. First, we use $\mathcal{I}$ to describe the subset of $2^{\mathcal{N}}$ that is compatible with all of the constraints imposed. If we restrict the set of vaccinated agents $V$ to belong to $\mathcal{I}$, which is part of a matroid $(\mathcal{Y},\mathcal{I})$, this constraint is called a matroid constraint.

\begin{definition}[Matroid]
Let $\mathcal{I}$ be a nonempty family of allowable subsets of $\mathcal{N}$. Then the tuple $(\mathcal{N},\mathcal{I})$ is a matroid if it satisfies$\colon$
\begin{itemize}
    \item  (Heredity) For any $D \subset E \subset \mathcal{N},$ if $E\in \mathcal{I}$, then $D\in \mathcal{I}$.
    \item (Augmentation) For any $D, E \in \mathcal{I}$, if $\vert D\vert < \vert E\vert$, then there exists an $x \in E\backslash D$ such that $D \cup \{x\} \in \mathcal{I}.$
\end{itemize}
\end{definition}

Let $\mathcal{N}_1$ and $\mathcal{N}_2$ be the disjoint subsets partitioned by $X_i$ ($\mathcal{N}_1\cup\mathcal{N}_2=\mathcal{N}$).
We can represent the targeting constraint by
\begin{equation}
    \mathcal{I}\equiv\{V\colon V\subseteq\mathcal{N}, \vert V\cap\mathcal{N}_1 \vert\leq d_1, \vert V\cap\mathcal{N}_2 \vert\leq d_2\}.
\end{equation}
We can show that this ($\mathcal{N},\mathcal{I}$) is a matroid referred to as a \textit{partition matroid}. First, we show heredity. For any $D\subset E,$ we must have $\vert D\cap\mathcal{N}_1\vert\leq \vert E\cap\mathcal{N}_1\vert$ and $\vert D\cap\mathcal{N}_2\vert\leq \vert E\cap\mathcal{N}_2\vert.$ If $E\in\mathcal{I},$ then it means $D$ must satisfy the targeting constraint in $\mathcal{I}.$ Next, for any $D, E \in \mathcal{I}$, we must have $\vert D\cap\mathcal{N}_1\vert, \vert E\cap\mathcal{N}_1\vert\leq d_1$ and $\vert D\cap\mathcal{N}_2\vert, \vert E\cap\mathcal{N}_2\vert\leq d_2.$ If  $\vert D\vert < \vert E\vert$, then either $\vert D\cap\mathcal{N}_1\vert< \vert E\cap\mathcal{N}_1\vert$ or $\vert D\cap\mathcal{N}_2\vert< \vert E\cap\mathcal{N}_2\vert$ or both. As a result, there must exist an element $x$ that belongs to $E\backslash D$ such that$\vert D\cup \{x\}\cap\mathcal{N}_1\vert\leq d_1$ and $\vert D\cup \{x\}\cap\mathcal{N}_2\vert\leq d_2.$

This problem of optimal treatment assignment subject to a partition matroid constraint is to maximize $F_n(V)$ over $V \in \mathcal{I}$. The following Algorithm 2 is guaranteed to produce a solution $\hat{V}' \in \mathcal{I}$. Greedy maximization algorithms subject to a partition matroid constraint performed for non-decreasing submodular functions attain at least $50\%$ of the optimal welfare.

\begin{spacing}{0.8}
\begin{algorithm}[H]
\begingroup
\renewcommand{\arraystretch}{1}
\linespread{1}
\SetAlgoLined
1$\colon$ \textbf{Input$\colon$} Dataset $\{S_{i},I_{i},R_{i},a_{i},b_{i}\}_{i=1}^{N},$ $\{A_{ij}\}_{i,j=1}^{N}$, estimated parameters $\{\hat{\beta}_{11},\hat{\beta}_{12},\hat{\beta}_{21},\hat{\beta}_{22},\hat{\gamma}_{1},\hat{\gamma}_{2}\}$, weight $\{g_i\}_{i=1}^N$, capacity constraint $d$, and targeting constraints $d_1, d_2$\;
\vskip 0.2 cm
2$\colon$ \textbf{Initialization}$\colon$ Starting from the empty set $V = \emptyset$ \;
\vskip 0.2 cm
\eIf {$\vert V\vert< d$}{
\vskip 0.2 cm
3$\colon$ \textbf{for} each $i \in \mathcal{N}\backslash V$ \textbf{do}\
\vskip 0.2 cm
4$\colon$ Compute the marginal gain $F_n(V+\{i\})\ -\ F_n(V)$\;
\vskip 0.2 cm
5$\colon$ Sort $i$ in order of decreasing marginal gain
\vskip 0.2 cm
6. \eIf { $\sum_{j\in V}a_j+a_{i(1)}\leq d_1\cap \sum_{j\in V}b_j+b_{i(1)}\leq d_2$ }{
\vskip 0.2 cm
7$\colon$ Add the 1st element of $i$ into $V$\;
}{
\vskip 0.2 cm
8$\colon$ Repeat step 6 with remaining $i$\;}
\vskip 0.2 cm
}{
\vskip 0.2 cm
\textbf{return} the set $V$\;
}
\caption{Targeting Constraint Greedy Algorithm}
\endgroup
\end{algorithm}
\end{spacing}
\bigskip
\begin{proposition}[\citealt{fisher1978analysis}] \label{prop.matroid}
Let $F_n : 2^{\mathcal{N}} \to \mathbb{R}$ be the simplified empirical welfare function as defined in (\ref{eq:26}) and $\hat{V}^{**} \in \arg \max_{V \in \mathcal{I}}F_n(V)$. The greedy maximization algorithm shown in Algorithm 2 outputs $\hat{V}' \in \mathcal{I}$ such that

\begin{equation}
    F_n(\hat{V}')\geq \frac{1}{2}F_n(\hat{V}^{**}).
\end{equation}
\end{proposition}

The performance guarantee of the greedy algorithm with targeting constraint is worse than the performance guarantee of Algorithm 1. This implies a trade-off between additional constraints and the accuracy of computation. In the next section, we discuss the welfare regret bounds of the allocation rules estimated by the above greedy algorithms.

\subsection{Perfect Treatment Assumption and Submodularity}\label{section72}
Recall Assumption \ref{assumpperfect} (Perfect Treatment)$\colon$ A vaccinated unit enters the Recovered state, regardless of its previous state (i.e., $\Pr(H_{1i}=R\vert v_i=1)=1$).
There are three possible ways to relax this assumption$\colon$
\begin{itemize}
    \item The recovered units can still spread disease.
    \item The recovered units will become susceptible after one period (few periods).
    \item Some percentage of vaccinated units remain susceptible or infected.
\end{itemize}
In the first case, if the person is recovered at $H_0$, she will spread the disease during the first period. In that case, the recovered neighbors of unit $i$ will be taken into account by the infection rate $q_i$. This will not, however, change the sign of our weighting matrix, hence submodularity (by Theorem \ref{proposition3.1}) still holds. In the second case, if unit $i$ is recovered in the first period (i.e., $H_{0i}=R$), she could become susceptible in the second period (i.e., $H_{1i}=S$). Then, she may be infected in the next period (i.e., $H_{2i}=I$). However, we only consider a one time period setting in this work, which rules out this risk. In the third case, varying this percentage only affects the coefficient of the linear term in the objective function (i.e., $\hat{c}_i$ in equation \ref{eq:23}), which is irrelevant to submodularity.

\section{Regret Bounds}
Following \citet{manski2004statistical} and the subsequent literature on statistical treatment rules, we use regret to evaluate the performance of our algorithm for vaccine allocation. Let $F: 2^{\mathcal{N}} \to \mathbb{R}$ be the population analogue of $F_n(\cdot)$ in (\ref{eq:26}), where the estimated parameters are replaced by the truth. The expected regret measures the average difference in the welfare between using the constrained optimal assignment rule $V^* \in \arg \max_{V \in \mathcal{V}_d} F(V)$ and using the constrained estimated greedy algorithm $\hat{V}$ obtained from Algorithm 1:
\begin{equation}
    F(V^*)-\E_{P^n}\big[F(\hat{V})\big]=\E_{P^n}\big[F(V^*)-F(\hat{V})\big]\geq 0,
\end{equation}
where $\E_{P^n}$ is the expectation with respect to the sampling uncertainty of the parameter estimates in the external studies. 

In this work, we assume that consistent estimators of effective contact rate and recovery rate are available from other studies. Generally, there is no requirement on the estimator except that Assumption \ref{ass5.1} needs to hold.

\begin{assumption}\label{ass5.1}
Let $\hat{\beta}_{sk}$ denote the estimate of effective contact rate between group $s$ and  group $k$, and $\hat{\gamma}_s$ denote the estimate of recovery rate in group $s$. The following properties need to hold:
\begin{equation} \label{eq:37}
    \mathbbm{P}\Big\{\abs{\hat{\beta}_{sk}-\beta_{sk}}\geq \epsilon\Big\}\leq 2e^{-2n\epsilon^2} 
    \quad \forall s,k = 1,2.
\end{equation}
\begin{equation} \label{eq:38}
    \mathbbm{P}\Big\{\abs{\hat{\gamma}_{s}-\gamma_{s}}\geq \epsilon\Big\}\leq 2e^{-2n\epsilon^2} 
    \quad \forall s = 1,2,
\end{equation}
where $\mathbbm{P}$ is the sampling distribution in another study that has sample size $n$.   
\end{assumption}

The above assumption is an exponential tail bound obtained by applying Hoeffding's large deviation inequality \citep{hoeffding1994probability}. Since $\beta_{sk}$ is the effective contact rate of the disease between group $s$ and $k$, and $\gamma_s$ is the recovery rate in group $s$, both are naturally bounded in $[0,1].$ Hence, common estimators (e.g., sample analog) meet the above condition. However, other tail bounds might apply for some other estimators, which do not necessarily have the same form as the above tail bound. Our approach can accommodate various tail bounds, such as the tail bound associated with the maximum likelihood estimator \citep{miao2010concentration}.

The estimators for the contact rates and recovery rates may come from different studies with different sample sizes. In this case, we can view $n$ in Assumption \ref{ass5.1} as the smallest sample size among the studies. 


In order to derive the uniform convergence rate of the welfare regret, we decompose regret into three components as follows.
\begin{equation}
 F(V^*)-F(\hat{V})=  \underbrace{F(V^*)-F_n(\hat{V}^*)}_{\circled{1}}+\underbrace{F_n(\hat{V}^*)-F_n(\hat{V})}_{\circled{2}}+\underbrace{F_n(\hat{V})-F(\hat{V})}_{\circled{3}},
\end{equation}
where $V^{\ast}$ is an oracle optimum $V^{\ast}= \arg\max_{V\in\mathcal{V}_d}F(V)$, $\hat{V}^*$ is a constrained optimal solution to the estimated welfare $\hat{V}^*=\arg\max_{V\in\mathcal{V}_d}F_n(V)$, and $\hat{V}$ is the output from the greedy maximization algorithm under the capacity constraint. Therefore, \circled{1} describes the regret we would attain if the constrained optimum could be computed exactly. \circled{2} measures the welfare loss introduced by the greedy algorithm. \circled{3} indicates the loss from using the estimated objective function instead of the true objective function. We compute the upper bound of each component separately and then combine them.

First, we start from the derivation of the upper bound of \circled{1}. This part is similar to the approach in \citet{kitagawa2018should}. Before looking at $V^*$, consider the following inequality, which holds for any $\widetilde{V}\in\mathcal{V}_d\colon$
\begin{equation}
    \begin{split}
        F(\widetilde{V})-F_n(\hat{V}^*)&\leq F(\widetilde{V})-F_n(\widetilde{V})\\
        &\big(\because F_n(\hat{V}^*)\geq F_n(\widetilde{V})\big)\\
        &\leq \sup_{V\in\mathcal{V}_d}\vert F_n(V)-F(V)\vert .
    \end{split}
\end{equation}
Since the above inequality applies to $F(\widetilde{V})$ for all $\widetilde{V},$ it also applies to $V^*\colon$
\begin{equation}
    F(V^*)-F_n(\hat{V}^*)\leq \sup_{V\in\mathcal{V}_d}\abs {F_n(V)-F(V)}.
\end{equation}
For the second component, we can obtain an upper bound by applying Theorem
\ref{theorem4.1}$\colon$
\begin{equation}\label{eq:41}
\begin{split}
    F_n(\hat{V}^*)-F_n(\hat{V})&\leq \frac{1}{e}F_n(\hat{V}^*)\\
                            &\leq \frac{1}{e}(F_n(\hat{V}^*)-F(\hat{V}^*))+\frac{1}{e}F(V^*)\\
                            &\leq \frac{1}{e}\abs{F_n(\hat{V}^*)-F(\hat{V}^*)}+\frac{1}{e}F(V^*)\\
                            &\leq \frac{1}{e}\sup_{V\in\mathcal{V}_d}\abs {F_n(V)-F(V)}+\frac{1}{e}F(V^*).
    \end{split}
\end{equation}
Similarly to the first component, the third component can be bounded as$\colon$
\begin{equation}
    F_n(\hat{V})-F(\hat{V})\leq\vert F_n(\hat{V})-F(\hat{V}) \vert\leq \sup_{V\in\mathcal{V}_d}\vert F_n(V)-F(V)\vert .
\end{equation}
Combining all the previous results, we obtain the upper bound of regret$\colon$
\begin{equation} \label{eq.Vhatregret}
\begin{split}
    F(V^*)-F(\hat{V})&\leq \left(2+\frac{1}{e} \right)\sup_{V\in\mathcal{V}_d}\abs{ F_n(V)-F(V)}+\frac{1}{e}F(V^*).
\end{split}
\end{equation}

Compared with the regret upper bound when one could compute $\hat{V}^{\ast}$, the regret upper bound shown in (\ref{eq.Vhatregret}) has one additional term $\frac{1}{e}\sup_{V\in\mathcal{V}_d}\abs {F_n(V)-F(V)}+\frac{1}{e}F(V^*)$. This additional term comes from equation (\ref{eq:41}) and captures the welfare loss induced by the use of greedy algorithm. As we characterize below, the first term converges to zero as $n \to \infty$ under Assumption \ref{ass5.1}, while the second term remains independent of the accuracy of the parameter estimates. A simulation study in Section 6 assesses the magnitude of the optimization error of the greedy algorithm, and shows numerically that the greedy algorithm yields an exact optimum for small network cases ($N=35$) at least. Based on this, we believe that the optimization error term of the greedy algorithm is much smaller than the universal theoretical bound $\frac{1}{e} F(V^{\ast})$. 

In the partition matroid (targeting constraint) case, by applying Proposition \ref{prop.matroid} and repeating the arguments to derive (\ref{eq.Vhatregret}), we obtain 
\begin{equation} \label{eq.Vhatregret2}
\begin{split}
    F(V^{**})-F(\hat{V}') & \leq \frac{5}{2}\sup_{V\in\mathcal{I}}\abs{ F_n(V)-F(V)}+\frac{1}{2}F(V^{**}),
\end{split}
\end{equation}
where $V^{**}$ is an oracle optimum under the targeting constraint, $V^{**} \in \arg \max_{V \in \mathcal{I}}F(V)$. 

In order to bound $\sup_{V\in\mathcal{V}_d}\abs{ F_n(V) - F(V)}$, we use the triangle inequality to find the bound of $\abs{F_n(V) - F(V)}\colon$
\begin{equation}
\footnotesize
\begin{split}
    \abs{ F_n(V) - F(V)} &= \abs{ \mathbf{v}^\intercal (\hat{W}-W) \mathbf{v} + (\hat{C}^\intercal-C^\intercal)\mathbf{v} - \mathbf{1}_{N\times 1}^\intercal (\hat{W}-W)\mathbf{v} - \mathbf{v}^\intercal (\hat{W}-W) \mathbf{1}_{N\times 1}}\\
                             &\leq \abs{ \mathbf{v}^\intercal (\hat{W}-W) \mathbf{v}} +\abs{ (\hat{C}^\intercal-C^\intercal)\mathbf{v}} +\abs{\mathbf{1}_{N\times 1}^\intercal (\hat{W}-W) \mathbf{v}}+\abs{\mathbf{v}^\intercal (\hat{W}-W) \mathbf{1}_{N\times 1}}\\
                             &\leq  \mathbf{v}^\intercal \abs{\hat{W}-W}\mathbf{v}+ \abs{(\hat{C}^\intercal-C^\intercal)}\mathbf{v}+ \mathbf{1}_{N\times 1}^\intercal \abs{\hat{W}-W} \mathbf{v}+\mathbf{v}^\intercal\abs{\hat{W}-W}\mathbf{1}_{N\times 1},
    \end{split}
\end{equation}
where the absolute value of a matrix or vector stands for the element-wise absolute values. Therefore, we can decompose the maximal deviation $\sup_{V\in\mathcal{V}}\abs{ F_n(V) - F(V)}$ into four parts$\colon$
\begin{equation} \label{eq:43}
\begin{split}
    \sup_{V\in\mathcal{V}_d}\abs{ F_n(V) - F(V)}&\leq \sup_{V\in\mathcal{V}_d} \mathbf{v}^\intercal \abs{\hat{W}-W} \mathbf{v}+\sup_{V\in\mathcal{V}_d} \abs{\hat{C}^\intercal-C^\intercal} \mathbf{v}\\ &+\sup_{V\in\mathcal{V}_d} \mathbf{1}_{N\times 1}^\intercal \abs{\hat{W}-W} \mathbf{v}+\sup_{V\in\mathcal{V}_d}\mathbf{v}^\intercal\abs{\hat{W}-W}\mathbf{1}_{N\times 1}.
    \end{split}
\end{equation}
Under Assumption \ref{ass5.1}, we can obtain an upper bound for the mean of each element in $\hat{W} - W$ and $\hat{C} - C$, as shown in the next lemma. 

\begin{lemma}\label{lamma5.1}
Under Assumption \ref{assumpundir},   \ref{assumpperfect}, and \ref{ass5.1}, we have
\begin{equation}
    \E_{P^n}\abs{\hat{w}_{ij}-w_{ij}}\leq\sqrt{\frac{1+\ln(2)}{2n}}\frac{A_{ij}g_i}{N},
    \hspace{4em}
    \E_{P^n}\abs{\hat{c}_i-c_i}\leq\sqrt{\frac{1+\ln(2)}{2n}}\frac{I_ig_i}{N}.
\end{equation}
\end{lemma}       

Combining this lemma with equations (\ref{eq.Vhatregret}) and (\ref{eq:43}), we obtain the following theorem:

\begin{theorem}\label{theorem5.1}
Let $N_M=\max_{i \in \mathcal{N}} \abs{N_i}$, $N_I$ be the total number of infected units, and $g=\max_{i \in \mathcal{N}} g_i$. Under Assumptions \ref{assumpundir}, \ref{assumpperfect}, and \ref{ass5.1}, we have
\begin{equation}
    \E_{P^n}\Big[F(V^*)-F(\hat{V})\Big]\leq \Bar{C} \cdot \frac{g\big[d\min\{N_M,d\}+2dN_M+\min\{N_I,d\}\big]}{N}\sqrt{\frac{1}{n}}+\frac{1}{e}F(V^*),
\end{equation}
where $\Bar{C}$ is a universal constant and $d$ is the number of available vaccine doses.
\end{theorem}
Proof of the above theorem is shown in the appendix. In Theorem \ref{theorem5.1}, we provided a distribution-free upper bound on the expected regret. We show that the convergence rate of the upper bound depends on the network data sample size $N$ and also the sample size $n$ for estimating the SIR parameters. At the same time, the regret upper bound is increasing in the complexity and the riskiness of the network. The intuition is that our algorithm finds it harder to identify the most valuable units when the maximum number of edges and the number of infected individuals in the network increases. The maximum individual weight $g$ also boosts the upper bound of regret. Moreover, our algorithm finds it harder to identify the best allocation rule when the number of possible combinations increase, which occurs when the capacity constraint is relaxed. This also implies the benefit of quarantine. Since quarantine controls the maximum number of connections in the network, the effectiveness of vaccine allocation is boosted by such government policy. Therefore, there is advantage to complementing a vaccine assignment policy with quarantine, which is evidenced by our simulation exercises.

\section{Simulation Exercises}
In this section, we use an Erd{\"{o}}s-Renyi model to generate random social networks. In each of the following tables, we use $100$ different networks and take the average of the outcome variable across all of the networks. We further show the standard deviation of in-sample welfare to understand the variation of network structure. We choose the probability of allocating a unit to group 1 to be $40\%$ and the probability of allocating a unit to group 2 to be $60\%$ (i.e., $\mathbbm{P}(X_i = G_1)=0.4$ and $\mathbbm{P}(X_i= G_2)=0.6$). In the epidemiological literature, researchers usually find the steady state of the SIR parameters. In order to identify the impact of varying the SIR parameters, we choose two different sets of parameter values to run the simulation. Throughout our simulation studies, we do not consider sampling errors in the parameter estimates and focus on optimizing welfare with the true parameter values plugged in. Table \ref{table 1} summarizes all the values of the SIR parameters that we have used. 

\begin{table}[H]
\begingroup
\setlength{\tabcolsep}{6pt} 
\renewcommand{\arraystretch}{1.1}
\linespread{1}
\footnotesize
\centering 
\begin{threeparttable}
 \begin{tabular}{@{}lcclcc@{}}
   \hline
   \toprule
   \textbf{\textit{Parameters}} &\textit{set} $1$&\textit{set} $2$ & \textbf{\textit{Parameters}}&\textit{set} $1$&\textit{set} $2$\\
    \midrule
   $\beta_{11}$ &\begin{tabular}[c]{@{}c@{}} $0.7$\end{tabular} &\begin{tabular}[c]{@{}c@{}} $0.8$\end{tabular}& $\beta_{12}$&\begin{tabular}[c]{@{}c@{}} $0.5$\end{tabular} &\begin{tabular}[c]{@{}c@{}} $0.5$\end{tabular} \\
   $\beta_{21}$ &\begin{tabular}[c]{@{}c@{}} $0.5$\end{tabular} &\begin{tabular}[c]{@{}c@{}} $0.7$\end{tabular} &$\beta_{22}$&\begin{tabular}[c]{@{}c@{}} $0.6$\end{tabular} &\begin{tabular}[c]{@{}c@{}} $0.7$\end{tabular}\\
   $\gamma_{1}$ &\begin{tabular}[c]{@{}c@{}} $0.1$\end{tabular} &\begin{tabular}[c]{@{}c@{}} $0.1$\end{tabular}
   &$\gamma_{2}$ &\begin{tabular}[c]{@{}c@{}} $0.05$\end{tabular} &\begin{tabular}[c]{@{}c@{}} $0.025$\end{tabular}\\
     \bottomrule
   \end{tabular}
   \end{threeparttable}
   \caption{Summary of the SIR parameter values}
\label{table 1}
\endgroup
\end{table}

In addition, we choose three different densities, $0.1$, $0.5$ and $1$, in order to identify the effect of network complexity. Here, $\textit{density} =1$ means that the network is fully connected (i.e., complete graph). We choose \textit{full} to understand the behaviour of our heuristic algorithm not only in the sparse network case but also in the densest case. We also compare three capacity constraints, $d =7\%N, 10\%N, 20\%N$, to evaluate the marginal performance gain of our greedy algorithm. We choose equal weight in the following comparisons. We, however, show the impact of changing weights on the number of vaccinated younger units in Table \ref{tablenumber}.


In the following sections, we compare our greedy algorithm with three familiar allocation rules. We first compare our algorithm with a brute force method in order to find the difference between the potentially sub-optimal greedy solution and the brute-force optimal solution. However, the number of possible combinations dramatically increases with the number of nodes and the capacity constraint. We cannot use a large number of agents to compute the brute force optimum in the simulation. Given this, in Section \ref{random}, we use a random assignment rule as a baseline to evaluate the performance of our algorithm in a large network setting. The third allocation rule that we compare our greedy algorithm with is an allocation rule which assigns the vaccine without considering network information. We compare the greedy algorithm with this third rule in Section \ref{no-net}.

\subsection{Comparing with Brute Force}\label{brute}
\begin{table}[H]
\begingroup
\setlength{\tabcolsep}{6pt} 
\renewcommand{\arraystretch}{0.88}
 \begin{adjustwidth}{0cm}{}
 \linespread{1}
\footnotesize
\centering 
\begin{threeparttable}
 \begin{tabular}{@{}lcccccc@{}}
   \hline
   \toprule
   \textbf{\textit{Allocation Rule}} & \multicolumn{3}{c}{\begin{tabular}[c]{@{}c@{}}\textit{Greedy Algorithm}\end{tabular}}& \multicolumn{3}{c}{\begin{tabular}[c]{@{}c@{}}\textit{Brute Force\footnote{We compare all possible combinations given the capacity constraint and select the set $V$ that maximizes $\mathcal{W}_n$.}}\end{tabular}}\\
  \cmidrule(lr){2-4}\cmidrule(lr){5-7}
  \textbf{\textit{Capacity Constraint}} & \begin{tabular}[c]{@{}c@{}}$d=7\%N$\end{tabular} & \begin{tabular}[c]{@{}c@{}}$d=10\%N$\end{tabular} & \begin{tabular}[c]{@{}c@{}}$d=20\%N$\end{tabular} & \begin{tabular}[c]{@{}c@{}}$d=7\%N$\end{tabular} & \begin{tabular}[c]{@{}c@{}}$d=10\%N$\end{tabular} & \begin{tabular}[c]{@{}c@{}}$d=20\%N$\end{tabular}\\
  \midrule
  \textbf{Parameter set 1}\\
 $N=500$, $density = 0.1$  &    0.60&  0.65& 0.77   &0.60&  0.65& 0.77   \\
                          &  (0.21)& (0.22)&(0.26)&(0.21)& (0.22)&(0.26)\\
  $N=500$, $density = 0.5$  &    0.47&  0.51& 0.63   &0.47&  0.51& 0.63\\
                          &(0.40) &(0.39)&(0.39)&(0.40) &(0.39)&(0.39)\\
  $N=500$, $density = 1$ &    0.33&  0.37& 0.49   & 0.33&  0.37& 0.49 \\
                          &(0.00)&(0.00)&(0.00)&(0.00)&(0.00)&(0.00)\\
  $N=800$, $density = 0.1$  &    0.58&  0.66& 0.76   &  0.58&  0.66& 0.76\\
                          &(0.23)  &(0.25)&(0.26)&(0.23)  &(0.25)&(0.26)\\
  $N=800$, $density = 0.5$  &    0.44&  0.52& 0.62   &0.44&  0.52& 0.62\\
                          &(0.38)  &(0.38)&(0.38)&(0.38)  &(0.38)&(0.38)\\
  $N=800$, $density = 1$ &    0.30&  0.36& 0.46   &  0.30&  0.36& 0.46\\
                          &(0.00)&(0.00)&(0.00)&(0.00)&(0.00)&(0.00)\\
  \textbf{Parameter set 2}\\
  $N=500$, $density = 0.1$  &    0.59&  0.64& 0.77   & 0.59&  0.64& 0.77\\
                            &(0.27)&(0.29)&(0.32)&(0.27)&(0.29)&(0.32)\\
  $N=500$, $density = 0.5$  &    0.42&  0.46& 0.59   &  0.42&  0.46& 0.59\\
                            &(0.49)&(0.49)&(0.49)&(0.49)&(0.49)&(0.49)\\
  $N=500$, $density = 1$    &    0.25&  0.29& 0.41   &  0.25&  0.29& 0.41  \\
                            &(0.00)&(0.00)&(0.00)&(0.00)&(0.00)&(0.00)\\
  $N=800$, $density = 0.1$  &    0.57&  0.65& 0.76   &   0.57&  0.65& 0.76\\
                          &(0.28)&(0.30)&(0.31)&(0.28)&(0.30)&(0.31)\\
  $N=800$, $density = 0.5$  &    0.40&  0.47& 0.58   & 0.40&  0.47& 0.58\\
                          &(0.47)&(0.47)&(0.47)&(0.47)&(0.47)&(0.47)\\
  $N=800$, $density = 1$ &    0.22&  0.29& 0.40   & 0.22&  0.29& 0.40  \\
  &(0.00)&(0.00)&(0.00)&(0.00)&(0.00)&(0.00)\\
  \bottomrule
   \end{tabular}
   \end{threeparttable}
   \caption{The value of welfare (the sum of probabilities of being healthy in the second period) averaged over 100 random networks (standard errors in parentheses). We use the Greedy Algorithm or the Brute Force algorithm to determine who in each network should be vaccinated.}
\label{table brute}
   \end{adjustwidth}
   \endgroup
\end{table}
\footnotetext[8]{We compare all possible combinations given the capacity constraint and select the set $V$ that maximizes $\mathcal{W}_n$.}

Since Theorem \ref{theorem4.1} shows the gap between the optimal solution and the heuristic result is at most $37\%,$ we want to explore this theoretical difference using numerical study. We list all the possible combinations and use brute force to search for the optimal solution given a manageable number of units. We specify the maximum number of units to be $N=35$, which is limited by computer performance. As the number of nodes increases, the possible number of combinations grows exponentially. The memory requirement and running time become impractical in a more realistic case. We recognize that the results from a small network may not be accurate in a large network setting, but help us to understand the regret of our greedy algorithm to some degree. We summarize the in-sample welfare $\mathcal{W}_n$ of these two approaches in Table \ref{table brute}.

In the small network case, we find that our greedy algorithm finds optimal allocation rules in all cases that we consider, which indicates a good performance of our method. We also notice that the welfare that is associated with the optimum decreases with the number of edges. As we relax the capacity constraint, welfare increases rapidly. The main purpose of this comparison is to get an idea of how much worse the empirical welfare at the greedy solution can be relative to the brute force optimum. More results are illustrated in the following two sections.
\subsection{Comparing With Random Assignment}\label{random}
In this section, we use a random assignment rule to define the baseline of vaccine allocation. We randomly draw an allocation $10,000$ times and calculate the average value of the outcome variable. Random allocation is one common assignment rule for policymakers. The purpose of this simulation is to learn about the improvement of our greedy allocation rule. In order to evaluate its performance in a relatively large network setting, we choose $N = 500$ and $800$. Table \ref{table random} records the main differences in terms of in-sample welfare between these two methods.

From Table \ref{table random}, we find that the performance of both methods decreases with the number of edges, which is also true for the first comparison. As the number of edges increase, the greedy algorithm finds it harder to identify who is relatively crucial in the network, which supports our interpretation of Theorem \ref{theorem5.1} in the previous section. This effect becomes more pronounced as the capacity constraint is relaxed. In the most extreme case, when everyone is connected with each other, the performance of our method is still better than the random assignment rule. This performance gap widens with the capacity constraint. We also find that the average welfare increases by $12\%$ when the capacity constraint increases by $0.1N$. Moreover, this improvement is robust with respect to the variation of number of nodes and the changes of density levels of network. The number of nodes decreases the performance of our method in a sparse network setting. For $N=800$, welfare in the densest network is $14\%$ lower than the welfare with $\textit{density}=0.5$, no matter which capacity constraint and parameter set we use.

\begin{table}[H]
\begingroup
\linespread{1}
\setlength{\tabcolsep}{6pt} 
\renewcommand{\arraystretch}{0.88}
 \begin{adjustwidth}{0cm}{}
\footnotesize
\centering 
\begin{threeparttable}
 \begin{tabular}{@{}lcccccc@{}}
   \hline
   \toprule
   \textbf{\textit{Allocation Rule}} & \multicolumn{3}{c}{\begin{tabular}[c]{@{}c@{}}\textit{Greedy Algorithm}\end{tabular}}& \multicolumn{3}{c}{\begin{tabular}[c]{@{}c@{}}\textit{Random Assignment\footnote{In Random allocation, We randomly select an assignment $10,000$ times for each network and take the average value of the outcome variable.}}\end{tabular}}\\
  \cmidrule(lr){2-4}\cmidrule(lr){5-7}
  \textbf{\textit{Capacity Constraint}} & \begin{tabular}[c]{@{}c@{}}$d=7\%N$\end{tabular} & \begin{tabular}[c]{@{}c@{}}$d=10\%N$\end{tabular} & \begin{tabular}[c]{@{}c@{}}$d=20\%N$\end{tabular} & \begin{tabular}[c]{@{}c@{}}$d=7\%N$\end{tabular} & \begin{tabular}[c]{@{}c@{}}$d=10\%N$\end{tabular} & \begin{tabular}[c]{@{}c@{}}$d=20\%N$\end{tabular}\\
  \midrule
  \textbf{Parameter set 1}\\
 $N=500$, $density = 0.1$  &    0.61&  0.65& 0.77   &0.57&    0.59& 0.66\\
                          &  (0.00)& (0.01)&(0.01)&(0.00)&(0.00)&(0.00)\\
  $N=500$, $density = 0.5$  &    0.61&  0.64& 0.76   &0.57&    0.59& 0.66\\
                          &(0.00) &(0.00)&(0.00)&(0.00)&(0.00)&(0.00)\\
  $N=500$, $density = 1$ &    0.61&  0.64& 0.76   &0.57&    0.59& 0.66\\
                          &(0.00)&(0.00)&(0.00)&(0.00)&(0.00)&(0.00)\\
  $N=800$, $density = 0.1$  &    0.59&  0.63& 0.75   &0.55&    0.57& 0.64\\
                          &(0.04)  &(0.04)&(0.04)&(0.04)&(0.04)&(0.04)\\
  $N=800$, $density = 0.5$  &    0.48&  0.51& 0.63   &0.44&    0.46& 0.53\\
                          &(0.07)  &(0.07)&(0.07)&(0.07)&(0.07)&(0.07)\\
  $N=800$, $density = 1$ &    0.34&  0.37& 0.49   &0.30&    0.32& 0.39\\
                          &(0.00)&(0.00)&(0.00)&(0.00)&(0.00)&(0.00)\\
  \textbf{Parameter set 2}\\
  $N=500$, $density = 0.1$  &    0.60&  0.64& 0.77   &0.56&    0.59& 0.66\\
                          &(0.01)&(0.01)&(0.01)&(0.00)&(0.00)&(0.00)\\
  $N=500$, $density = 0.5$  &    0.60&  0.64& 0.76   &0.56&    0.59& 0.66\\
                          &(0.00)&(0.00)&(0.00)&(0.00)&(0.00)&(0.00)\\
  $N=500$, $density = 1$ &    0.60&  0.64& 0.76      &0.56&    0.59& 0.66\\
                          &(0.00)&(0.00)&(0.00)&(0.00)&(0.00)&(0.00)\\
  $N=800$, $density = 0.1$  &    0.58&  0.62& 0.74   &0.54&    0.56& 0.63\\
                          &(0.05)&(0.05)&(0.05)&(0.05)&(0.05)&(0.05)\\
  $N=800$, $density = 0.5$  &    0.44&  0.48& 0.60   &0.40&    0.43& 0.50\\
                          &(0.08)&(0.08)&(0.08)&(0.08)&(0.08)&(0.08)\\
  $N=800$, $density = 1$ &    0.26&  0.30& 0.42   &0.23&    0.25& 0.32\\
  &(0.00)&(0.00)&(0.00)&(0.00)&(0.00)&(0.00)\\
  \bottomrule
   \end{tabular}
   \end{threeparttable}
   \caption{The value of welfare (the sum of probabilities of being healthy in the second period) averaged over 100 random networks (standard errors in parentheses). We use the Greedy Algorithm or the Random allocation to determine who in each network should be vaccinated.}
\label{table random}
     \end{adjustwidth}
   \endgroup
\end{table}
\footnotetext[9]{In Random allocation, We randomly select an assignment $10,000$ times for each network and take the average value of the outcome variable.}

\begin{figure}[H]
\centering
\begin{minipage}[t]{0.48\textwidth}
\centering
\includegraphics[width=6.5cm,height=6.5cm]{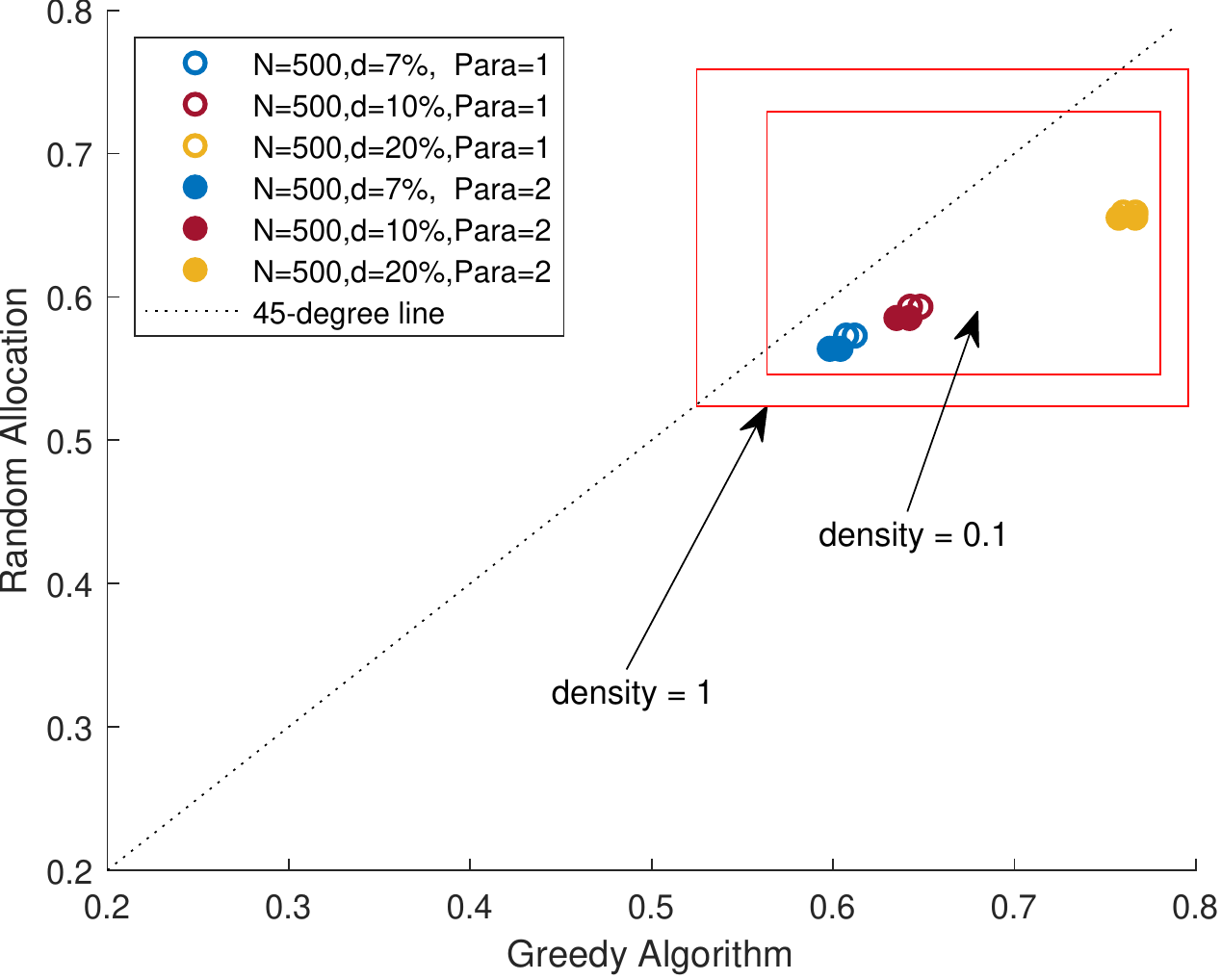}
\end{minipage}
\begin{minipage}[t]{0.48\textwidth}
\centering
\includegraphics[width=6.5cm,height=6.5cm]{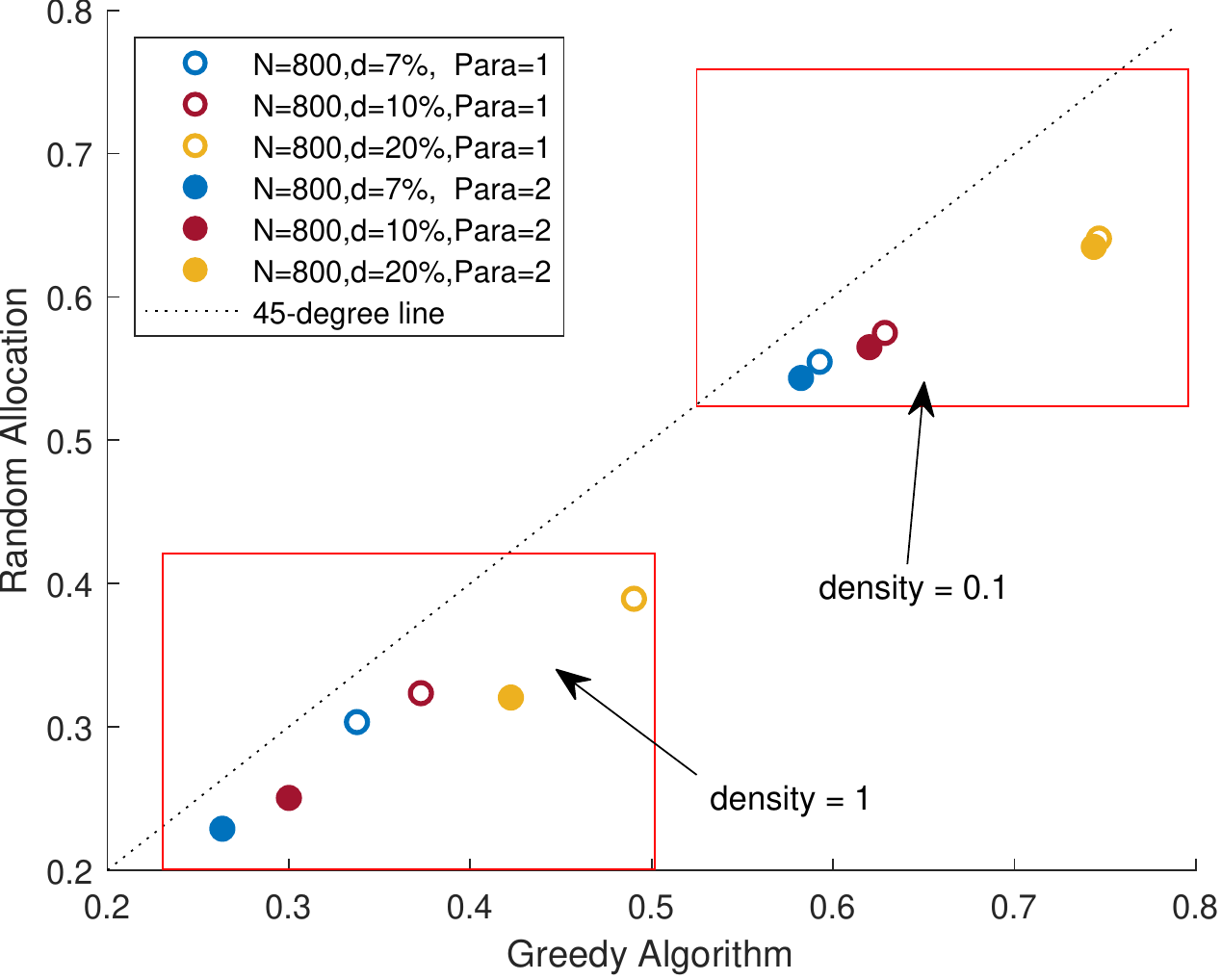}
\end{minipage}
\caption{Comparison between Greedy Algorithm and Random allocation}
\label{figure1}
\end{figure}

If we look at the random assignment rule in Table \ref{table random}, its performance is much worse than the performance of the greedy algorithm. This difference increases when the complexity of and the number of nodes in the network increase. The performance of the random assignment rule improves as we relax the capacity constraint. However, this improvement is only about $7\%$ when the capacity constraint increases by $0.1N.$ Compared with the greedy algorithm, random assignment is less effective. Given its scarcity, we waste considerable resources by randomly assigning the vaccine. Looking at the situation of full edges, the performance of random allocation is inferior. The ratio of the welfare attained by random allocation to the welfare attained by the greedy algorithm is illustrated in Figure \ref{figure1}. This ratio increases slowly with the number of edges and deceases with the number of nodes in the network. In addition, the ratio decreases in an obvious way with the number of vaccines that are available.

\subsection{Comparing With Targeting Without Network Information}\label{no-net}

\begin{table}[H]
\begingroup
\setlength{\tabcolsep}{6pt} 
\renewcommand{\arraystretch}{0.88}
\linespread{1}
 \begin{adjustwidth}{0cm}{}
\footnotesize
\centering 
\begin{threeparttable}
 \begin{tabular}{@{}lcccccc@{}}
   \hline
   \toprule
   \textbf{\textit{Allocation Rule}} & \multicolumn{3}{c}{\begin{tabular}[c]{@{}c@{}}\textit{Greedy Algorithm}\end{tabular}}& \multicolumn{3}{c}{\begin{tabular}[c]{@{}c@{}}\textit{TWNI\footnote{We assign the vaccine only to the second group (i.e., only older people receive the vaccine).}}\end{tabular}}\\
  \cmidrule(lr){2-4}\cmidrule(lr){5-7}
  \textbf{\textit{Capacity Constraint}} & \begin{tabular}[c]{@{}c@{}}$d=7\%N$\end{tabular} & \begin{tabular}[c]{@{}c@{}}$d=10\%N$\end{tabular} & \begin{tabular}[c]{@{}c@{}}$d=20\%N$\end{tabular} & \begin{tabular}[c]{@{}c@{}}$d=7\%N$\end{tabular} & \begin{tabular}[c]{@{}c@{}}$d=10\%N$\end{tabular} & \begin{tabular}[c]{@{}c@{}}$d=20\%N$\end{tabular}\\
  \midrule
  \textbf{Parameter set 1}\\
 $N=500$, $density = 0.1$  &    0.61&  0.65& 0.77   &0.57&    0.59& 0.65\\
                          &  (0.00)& (0.01)&(0.01)&(0.00)&(0.01)&(0.01)\\
  $N=500$, $density = 0.5$  &    0.61&  0.64& 0.76   &0.57&    0.59& 0.65\\
                          &(0.00) &(0.00)&(0.00)&(0.00)&(0.00)&(0.00)\\
  $N=500$, $density = 1$ &    0.61&  0.64& 0.76   &0.57&    0.59& 0.65\\
                          &(0.00)&(0.00)&(0.00)&(0.00)&(0.00)&(0.00)\\
  $N=800$, $density = 0.1$  &    0.59&  0.63& 0.75   &0.56&    0.58& 0.65\\
                          &(0.04)  &(0.04)&(0.04)&(0.04)&(0.04)&(0.04)\\
  $N=800$, $density = 0.5$  &    0.48&  0.51& 0.63   &0.44&    0.47& 0.54\\
                          &(0.07)  &(0.07)&(0.07)&(0.07)&(0.07)&(0.07)\\
  $N=800$, $density = 1$ &    0.34&  0.37& 0.49   &0.30&    0.33& 0.40\\
                          &(0.00)&(0.00)&(0.00)&(0.00)&(0.00)&(0.00)\\
  \textbf{Parameter set 2}\\
  $N=500$, $density = 0.1$  &    0.60&  0.64& 0.77   &0.56&    0.58& 0.65\\
                          &(0.01)&(0.01)&(0.01)&(0.01)&(0.01)&(0.01)\\
  $N=500$, $density = 0.5$  &    0.60&  0.64& 0.76   &0.56&    0.58& 0.65\\
                          &(0.00)&(0.00)&(0.00)&(0.00)&(0.00)&(0.00)\\
  $N=500$, $density = 1$ &    0.60&  0.64& 0.76   &0.56&    0.58& 0.65\\
                          &(0.00)&(0.00)&(0.00)&(0.00)&(0.00)&(0.00)\\
  $N=800$, $density = 0.1$  &    0.58&  0.62& 0.74   &0.55&    0.57& 0.64\\
                          &(0.05)&(0.05)&(0.05)&(0.05)&(0.05)&(0.05)\\
  $N=800$, $density = 0.5$  &    0.44&  0.48& 0.60   &0.41&    0.43& 0.50\\
                          &(0.08)&(0.08)&(0.08)&(0.08)&(0.08)&(0.08)\\
  $N=800$, $density = 1$ &    0.26&  0.30& 0.42   &0.23&    0.26& 0.33\\
  &(0.00)&(0.00)&(0.00)&(0.00)&(0.00)&(0.00)\\
  \bottomrule
   \end{tabular}
   \end{threeparttable}
   \caption{The value of welfare (the sum of probabilities of being healthy in the second period) averaged over 100 random networks (standard errors in parentheses). We use the Greedy Algorithm or the Targeting Without Network Information allocation to determine who in each network should be vaccinated}
\label{table no-net}
   \end{adjustwidth}
   \endgroup
\end{table}

\footnotetext[10]{We assign the vaccine only to the second group (i.e., only older people receive the vaccine).}

Usually, in the literature on treatment assignment, researchers use observational data or experimental data without network structure information to study the optimal policy. As a result, the allocation regime assigns the treatment without considering spillover effects, which could lead to a sub-optimal result. We call this kind of regime Targeting Without Network Information (\textbf{TWNI}). In this simulation, we want to learn the welfare loss from using TWNI versus our method.

Generally, TWNI assigns treatment based on personal characteristics. In this study, we only have one covariate$\colon$ \textit{age}. This means either the old group receives the vaccine or the young group receives the vaccine. Under the previous setting (i.e., older people are more likely to be infected and to die), group 2 will consume the entire vaccine allocation. Given different capacity constraints, this assignment rule selects units to be vaccinated from group 2 until the upper bound is reached. Table \ref{table no-net} indicates the results for TWNI allocation are similar to those for random allocation. In addition, despite the outcome value varying with the SIR parameters, the sizable improvement from using network information to allocate vaccination is quite robust to variations in the size and density of network.

\begin{figure}[H]
\centering
\begin{minipage}[t]{0.48\textwidth}
\centering
\includegraphics[width=6.5cm,height=6.5cm]{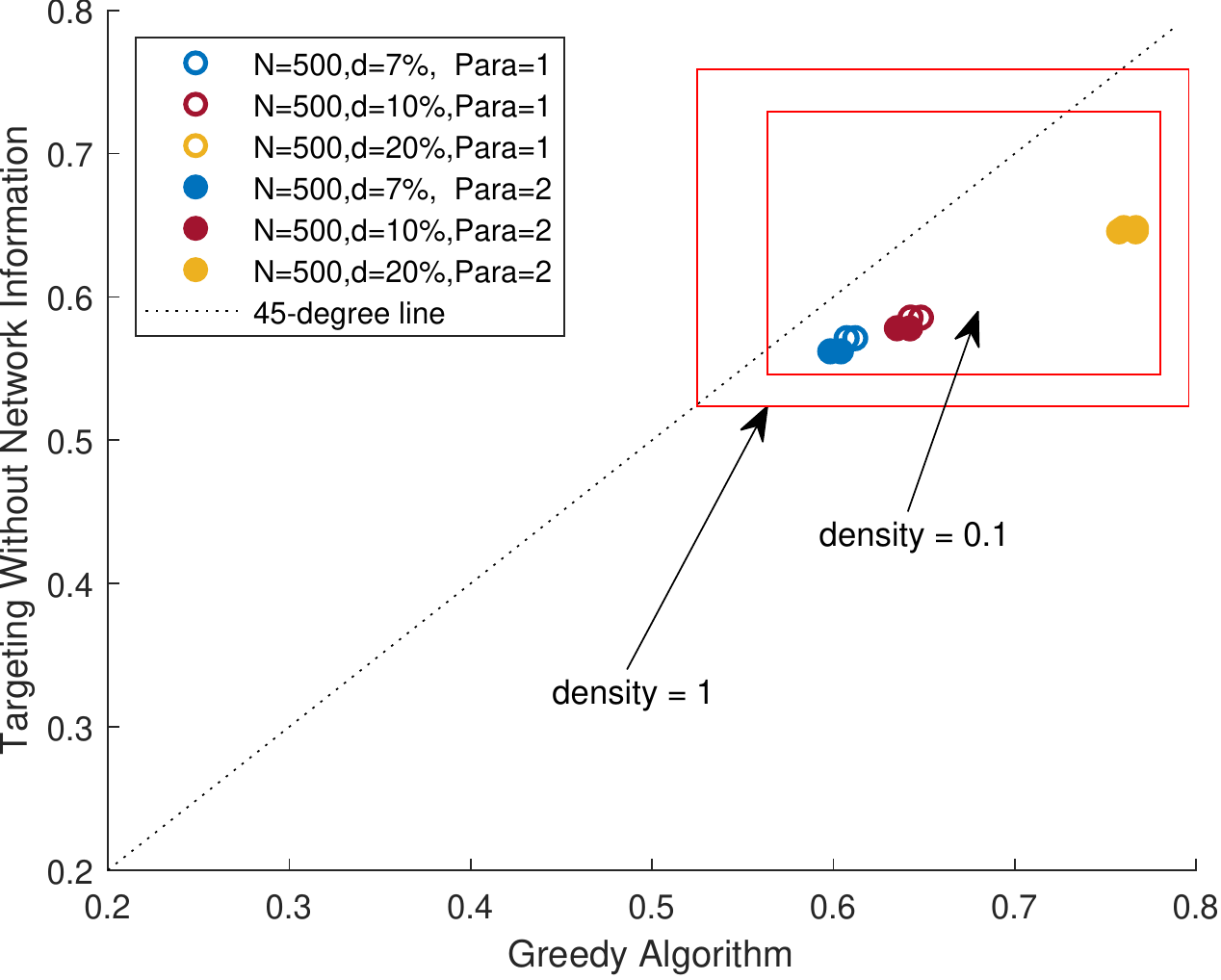}
\end{minipage}
\begin{minipage}[t]{0.48\textwidth}
\centering
\includegraphics[width=6.5cm,height=6.5cm]{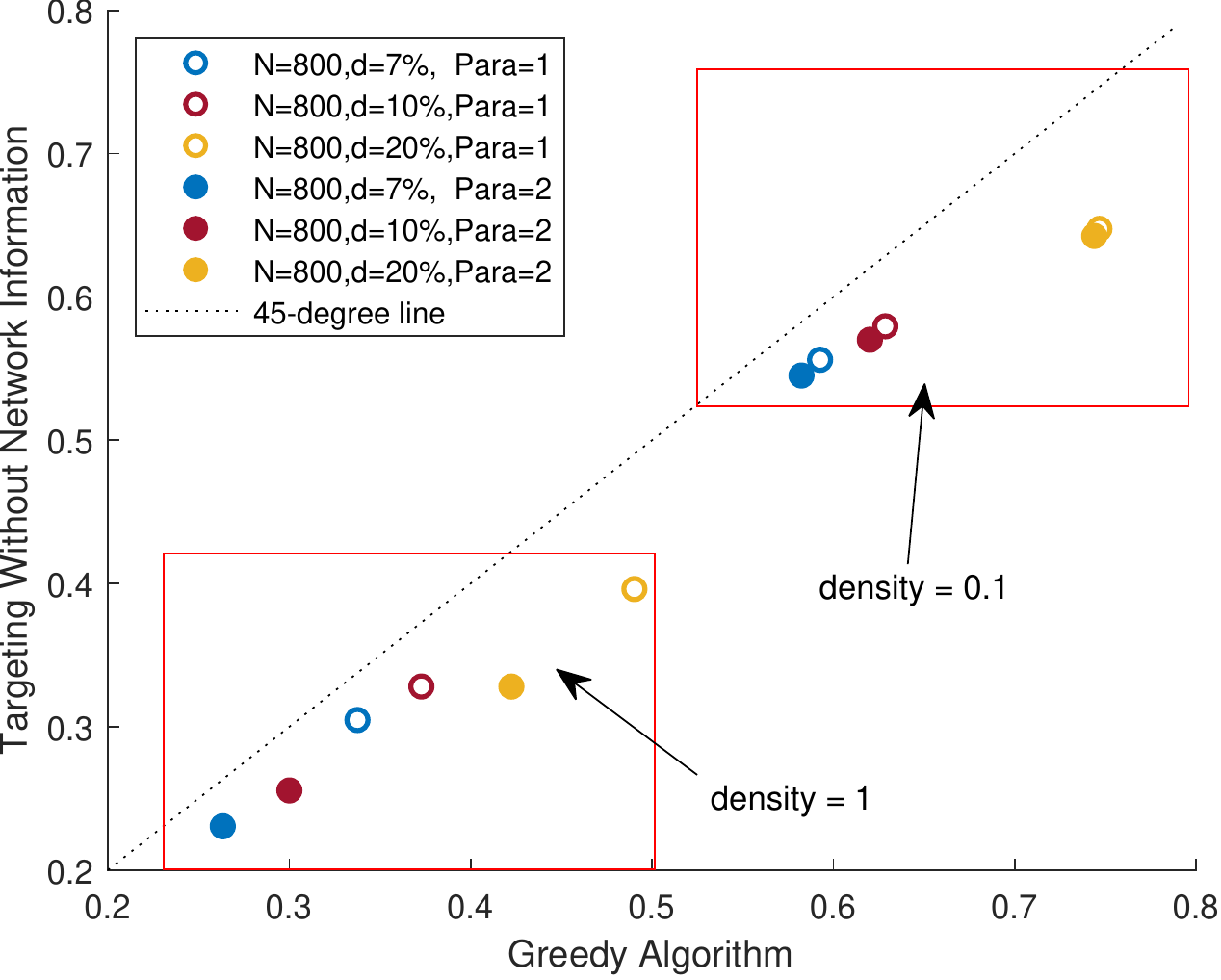}
\end{minipage}
\caption{Comparison between Greedy Algorithm and Targeting Without Network Information}
\label{figure2}
\end{figure}

Our numerical study shows that if the number of available vaccine doses is small, the loss from ignoring network information is relatively small too (around $4\%$). This loss increases dramatically, however, with the number of available vaccines. In addition, the performance gap between our greedy algorithm and the other two allocation methods decreases with the network complexity (i.e., the number of edges). Under what might be described as a lockdown policy, the density of the network is maintained at a relatively low level, which raises the cost of ignoring spillovers. This cost also increases with the number of units in the population, which is a problem in a more realistic setting. The performance improvement from considering network information is robust to variation of the SIR parameters, and an allocation rule which ignores spillovers waste a sizeable proportion of a scarce resource.

In Table \ref{tablenumber}, we illustrate the impact on the percentage of vaccinated younger units by varying the weight choice $g_i$ (In this simulation exercise, we choose equal weight for the units in same group). If we assign weight $g_1=1.5$ for $G_1$, we find all the vaccines are consumed by younger units. Comparing with the equal weight case, this number changes dramatically. Moreover, we find our greedy algorithm offers more vaccines to younger units in the case of parameter set 2 than parameter set 1, i.e., when the transmission rate parameters are higher within and across the groups.

\begin{table}[H]
\begingroup
\setlength{\tabcolsep}{6pt} 
\renewcommand{\arraystretch}{1}
\linespread{1}
 \begin{adjustwidth}{0cm}{}
\footnotesize
\centering 
\begin{threeparttable}
 \begin{tabular}{@{}lcccccc@{}}
   \hline
   \toprule
   \textbf{\textit{Weight Choice}} & \multicolumn{2}{c}{\begin{tabular}[c]{@{}c@{}}\textit{Weight} $g_1=1$, $g_2=1$\end{tabular}}&\multicolumn{2}{c}{\begin{tabular}[c]{@{}c@{}}\textit{Weight} $g_1=1.1$, $g_2=1$\end{tabular}}&\multicolumn{2}{c}{\begin{tabular}[c]{@{}c@{}}\textit{Weight} $g_1=1.5$, $g_2=1$\end{tabular}}\\
  \cmidrule(lr){2-3}\cmidrule(lr){4-5}\cmidrule(lr){6-7}
  \textbf{\textit{Capacity Constraint}} & \begin{tabular}[c]{@{}c@{}}$d=7\%N$\end{tabular} & \begin{tabular}[c]{@{}c@{}}$d=20\%N$\end{tabular} & \begin{tabular}[c]{@{}c@{}}$d=7\%N$\end{tabular} & \begin{tabular}[c]{@{}c@{}}$d=20\%N$\end{tabular} & \begin{tabular}[c]{@{}c@{}}$d=7\%N$\end{tabular} & \begin{tabular}[c]{@{}c@{}}$d=20\%N$\end{tabular}\\
  \midrule
  \textbf{Parameter set 1}\\
   $N=500$, $density = 0.1$&    9\%&  17\% & 80\% &  61\%  &100\%&     100\%\\
   $N=500$, $density = 0.5$&    0\%&  1\%  & 100\%&  94\%   &100\%&    100\%\\
   $N=500$, $density = 1$&      9\%&  17\%  & 80\%&  61\%   &100\%&    100\%\\
   $N=800$, $density = 0.1$&    11\%&  14\% & 84\%&  69\%   &100\%&    100\%\\
   $N=800$, $density = 0.5$&    0\%&  1\%  & 100\%&   95\% &100\%&    100\%\\
   $N=800$, $density = 1$&      0\%&   0\% & 100\%&   100\%  &100\%&     100\%\\
   \textbf{Parameter set 2}\\
   $N=500$, $density = 0.1$&    31\%&  27\%&   89\%& 64\%&      100\%& 100\%\\
   $N=500$, $density = 0.5$&    3\%&   13\%&   100\%& 95\%&       100\%& 100\%\\
   $N=500$, $density = 1$&      31\%&   27\%&   89\%& 64\%&       100\%& 100\%\\
   $N=800$, $density = 0.1$&    23\%&  26\%&    91\%& 71\%&      100\%& 100\%\\
   $N=800$, $density = 0.5$&    5\%&   10\%&    100\%& 98\%&       100\%& 100\%\\
   $N=800$, $density = 1$&      0\%&   0\%&    100\%& 100\%&        100\%& 100\%\\
 \bottomrule
   \end{tabular}
   \end{threeparttable}
   \caption{The percentage of vaccinated younger units in the second period under the vaccine allocation policies obtained by Greedy Algorithm, averaged over 100 random networks. We choose three different sets of weights in this comparison}
\label{tablenumber}
   \end{adjustwidth}
   \endgroup
\end{table}


\section{Conclusion}
In this work, we have introduced a novel method to estimate individualized vaccine allocation rules under network interference. We introduce the heterogeneous-interacted-SIR model to specify the spillover effects of infectious disease. We show that the welfare objective function of the vaccine allocation problem is non-decreasing and submodular, and so is its empirical analogue formed by plugging in the estimates of the SIR parameters. Based on this specific diminishing returns property, we provide a greedy algorithm with performance guarantee under two different exogenous constraints, which can easily accommodate various targets that policymakers commonly face in reality. Moreover, we show that this algorithm implies an upper bound for regret that converges uniformly at $\mathcal{O}(n^{-1/2})$. Using simulation, we point out the importance of considering network information in the allocation problem. 

Several open questions and extensions are worth considering in future work. First, this paper considered a one-time vaccine allocation. We did not consider if there are multiple allocation periods, and how to decide the allocation dynamically. A relevant important question is how to jointly optimize allocations and timing of first- and second-doses of vaccines, as recently discussed for Covid-19 vaccines in \citet{maier2021potential}, \citet{tuite2021alternative}, and \citet{wang2021impacts}). Moreover, we do not study how the vaccine allocation rule impacts on the outcome variables after multiple periods. As discussed in \citet{bu2020likelihood}, changes to the network structure should be considered in a dynamic setting. Second, we only compare the greedy algorithm with the brute-force optimum in a small network. Other than the universal bounds of Theorem \ref{theorem5.1}, we do not know the performance of our method relative to the optimal solution in the large network data setting. 
Third, we did not impose any other constraints than the capacity constraint and the targeting constraint. For interpretability and fairness, we may want to additionally restrict the policy rule as a simple function of observed covariates. We regard these as interesting questions that are worthy of consideration.

\bibliographystyle{ecta}
\bibliography{ref}
\begin{appendices}

\section{The Transmission Term}
Consider a susceptible individual $i$ with $\kappa_s$ contacts which depends on his own characteristics at each period. Of these contacts, a fraction $\sum_{j\in N_i}I_j(1-v_j)a_j/\abs{N_i}$ are contacts with infected neighbors from group 1, and a fraction $\sum_{j\in N_i}I_j(1-v_j)b_j/\abs{N_i}$ are contacts with infected neighbors from group 2. If we define $c_{ij}$ as the probability of successful disease transmission at each contact, then $1 -c_{sk} $ is the probability that transmission between group $s$ and group $k$ does not take place. Therefore, we have the probability that a unit $i$ is not infected in one time period$\colon$
\begin{equation}
\begin{split}
    1- q_i &=(1-c_{11})^{\frac{\kappa_1\sum_{j\in N_i}I_j(1-v_j)a_ja_i}{\abs{N_i}}}\cdot(1-c_{12})^{\frac{\kappa_1\sum_{j\in N_i}I_j(1-v_j)b_ja_i}{\abs{N_i}}}\\
    &\quad\cdot(1-c_{21})^{\frac{\kappa_2\sum_{j\in N_i}I_j(1-v_j)a_jb_i}{\abs{N_i}}} \cdot(1-c_{22})^{\frac{\kappa_2\sum_{j\in N_i}I_j(1-v_j)b_jb_i}{\abs{N_i}}}.\\
\end{split}
\end{equation}
We now define $\beta_{sk}=-\kappa_{s}\ln(1-c_{sk})$ and plug it into the expression for $1-q_i$, which allows us to rewrite the above equation as$\colon$
\begin{equation}
        q_i = 1 - e^{-z},\\
\end{equation}
where
\begin{equation}
\begin{split}
    z &= \frac{\beta_{11}}{\abs{N_i}}\sum_{j\in N_i}I_j(1-v_j)a_ja_i+\frac{\beta_{12}}{\abs{N_i}}\sum_{j\in N_i}I_j(1-v_j)b_j a_i\\
    &\quad+\frac{\beta_{21}}{\abs{N_i}}\sum_{j\in N_i}I_j(1-v_j)a_j b_i+\frac{\beta_{22}}{\abs{N_i}}\sum_{j\in N_i}I_j(1-v_j)b_j b_i.\\
\end{split}
\end{equation}
Recalling that $e^x= 1+x+\frac{x^2}{2!}+\frac{x^3}{3!}+\cdots,$ we now have the probability of infection at each time period is
\begin{equation}
    q_i \simeq z.
\end{equation}

\section{Lemmas}

\subsection{Preliminary Lemma}
In this section we collect a set of lemmas from past literature that we use in our proofs.

\begin{lemma}[Proposition 6.3 \citet{bach2011learning}]\label{lemmabach}
Let $Q\in\mathbbm{R}^{p\times p}$, $q\in \mathbbm{R}^{p}$, and $\mathcal{N} = \{1,2, \dots, p \}$. For $A \in 2^{\mathcal{N}}$, define $1_{A} = (1_{1 \in A}, \dots, 1_{p \in A} )'$. The function $F\colon A\mapsto q^{\intercal}1_A+\frac{1}{2}1_A^{\intercal}Q1_{A}$ is submodular if and only if all off-diagonal elements of Q are non-positive.
\end{lemma}

\begin{lemma}[Theorem 2.2 \citet{cunningham1985minimum}]\label{lemmacunn}
Function $F$ is a cut function if and only if$\colon$ For any three disjoint subsets $A,$ $B,$ $C$ of S,
\begin{equation*}
    F(A\cup B\cup C) = F(A\cup B)+F(A\cup C)+F(B\cup C)-F(A)-F(B)-F(C)+F(\emptyset).
\end{equation*}
\end{lemma}

\quad The following lemmas are some common techniques that are often used in the statistical learning literature, as reviewed in \citet{lugosi2002pattern}.

\begin{lemma}[Hoeffding's inequality \citet{hoeffding1994probability}]\label{lemmahoeff}
Let $X_1,...,X_n$ be independent bounded random variables such that $X_i$ falls in the interval $[a_i,b_i]$ with probability one. Denote their sum by $S_n=\sum_{i=1}^n X_i$. Then for any $\epsilon>0$ we have
\begin{equation*}
    \mathbbm{P}\{S_n-\E S_n\geq \epsilon\}\leq e^{-2e^2/\sum_{i=1}^n (b_i-a_i)^2},
\end{equation*}
and
\begin{equation*}
    \mathbbm{P}\{S_n-\E S_n\leq -\epsilon\}\leq e^{-2e^2/\sum_{i=1}^n (b_i-a_i)^2}.
\end{equation*}
\end{lemma}

\subsection{Proof of Lemma \ref{proposition3.1}}
Let $\hat{W}\in\mathbb{R}^{N\times N}$ and $\hat{C}\in\mathbb{R}^{N}.$ Then the function $F_n\colon V \mapsto\mathbf{v} ^\intercal \hat{W} \mathbf{v} + \hat{C}^\intercal \mathbf{v} - \mathbf{1}_{N\times 1}^\intercal \hat{W} \mathbf{v} - \mathbf{v} ^\intercal \hat{W} \mathbf{1}_{N\times 1}$ is submodular if and only if $\hat{w}_{ij}\leq 0\ \forall i\neq j.$
\begin{proof}
The first step is to show our objective function is a cut function based on Lemma \ref{lemmacunn}. In our case, simply consider three arbitrary disjoint sets $A, B, C\subseteq \mathcal{N}$.

\begin{equation}
    F_n(A)=\sum_{j\in A}\{\hat{w}_{jj}+\hat{c}_j\}+\sum_{i\neq j\in A}\{\hat{w}_{ij}\}-\sum_{j\in A}\sum_{m=1}^{N}\{\hat{w}_{mj}+\hat{w}_{jm}\},
\end{equation}
\begin{equation}
    F_n(B)=\sum_{j\in B}\{\hat{w}_{jj}+\hat{c}_j\}+\sum_{i\neq j\in B}\{\hat{w}_{ij}\}-\sum_{j\in B}\sum_{m=1}^{N}\{\hat{w}_{mj}+\hat{w}_{jm}\},
\end{equation}

\begin{equation}
\begin{split}
    F_n(\{A\cup B\})&=\sum_{j\in \{A\cup B\}}\{\hat{w}_{jj}+\hat{c}_j\}+\sum_{i\neq j\in \{A\cup B\}}\{\hat{w}_{ij}\}-\sum_{j\in \{A\cup B\}}\sum_{m=1}^{N}\{\hat{w}_{mj}+\hat{w}_{jm}\}\\
    &=\sum_{j\in A }\{\hat{w}_{jj}+\hat{c}_j\}+\sum_{i\neq j\in A}\{\hat{w}_{ij}\}-\sum_{j\in A}\sum_{m=1}^{N}\{\hat{w}_{mj}+\hat{w}_{jm}\}\\&+\sum_{j\in B}\{\hat{w}_{jj}+\hat{c}_j\}+\sum_{i\neq j\in B}\{\hat{w}_{ij}\}-\sum_{j\in B}\sum_{m=1}^{N}\{\hat{w}_{mj}+\hat{w}_{jm}\}\\
    &=F_n(A)+F_n(B).
    \end{split}
\end{equation}
Therefore, we have$\colon$
\begin{equation}
    \begin{split}
            F_n(\{A\cup B\}\cup C)&=F_n(\{A\cup B\})+F_n(C)\\
    &=F_n(A)+F_n(B)+F_n(C)
    \end{split}
\end{equation}
Combining the previous results, we get$\colon$
\begin{equation}
    \begin{split}
        F_n(\{A\cup B\}\cup C) &= F_n(A\cup B)+F_n(A\cup C)+F_n(B\cup C)\\
        &\quad-F_n(A)-F_n(B)-F_n(C)-F_n(\emptyset),
    \end{split}
\end{equation}
since $F_n(\emptyset)=0.$

Now, we have shown that $F_n(V)$ is a cut function. The next step is to find the sufficient and necessary conditions for submodularity of the cut function. Lemma \ref{lemmabach} indicates, for any cut function which can be written as a quadratic function plus a linear part, submodularity holds if and only if all off-diagonal elements of the weighting matrix are non-positive. That requires $\hat{w}_{ij}\leq 0,\ \forall i\neq j.$ 
\end{proof}

\subsection{Proof of Lemma \ref{lamma5.1}}
Under Assumption \ref{assumpundir},   \ref{assumpperfect}, and \ref{ass5.1}, we have
\begin{equation}
    \E_{P^n}\abs{\hat{w}_{ij}-w_{ij}}\leq\sqrt{\frac{1+\ln(2)}{2n}}\frac{A_{ij}g_i}{N},
    \hspace{4em}
    \E_{P^n}\abs{\hat{c}_i-c_i}\leq\sqrt{\frac{1+\ln(2)}{2n}}\frac{I_ig_i}{N}.
\end{equation}
\begin{proof} 
We first prove the upper bound of $\E_{P^n}\abs{\hat{w}_{ij}-w_{ij}}.$
\begin{equation}
\begin{split}
    \hat{w}_{ij}-w_{ij}&=\frac{S_i g_iA_{ij}I_j}{\abs{N_i}N}\Big[(\beta_{11}-\hat{\beta}_{11})a_ia_j+ (\beta_{12}-\hat{\beta}_{12})a_ib_j +(\beta_{21}-\hat{\beta}_{21}) b_ia_j +(\beta_{22}-\hat{\beta}_{22})b_ib_j\Big].
        \end{split}
\end{equation}
If we take the absolute value and expectation of each side, by the triangle inequality, we get 
\begin{equation}\label{eq:47}
    \begin{split}
   \E_{P^n}\abs{\hat{w}_{ij}-w_{ij}} &= \E_{P^n}\left\vert\frac{S_i g_iA_{ij}I_j}{\abs{N_i}N}\Big[(\hat{\beta}_{11}-\beta_{11})a_ia_j+ (\hat{\beta}_{12}-\beta_{12})a_ib_j\right.\\
   &\quad\left.+(\hat{\beta}_{21}-\beta_{21}) b_ia_j +(\hat{\beta}_{22}-\beta_{22})b_ib_j\Big]\right\vert\\
   &\leq \frac{S_i g_iA_{ij}I_ja_ia_j}{\abs{N_i}N}\E_{P^n}\abs{\hat{\beta}_{11}-\beta_{11}}+ \frac{S_i g_iA_{ij}I_ja_ib_j}{\abs{N_i}N}\E_{P^n}\abs{\hat{\beta}_{12}-\beta_{12}} \\&\quad+\frac{S_i g_iA_{ij}I_jb_ia_j}{\abs{N_i}N}\E_{P^n}\abs{\hat{\beta}_{21}-\beta_{21}} +\frac{S_i g_iA_{ij}I_jb_ib_j}{\abs{N_i}N}\E_{P^n}\abs{\hat{\beta}_{22}-\beta_{22}}.
    \end{split}
\end{equation}
Since $\beta_{sk}$ is the effective contact rate of the disease between group $s$ and $k$, it is naturally bounded in $[0,1].$ We can apply Lemma \ref{lemmahoeff} to get the upper bound of each component$\colon$
\begin{equation}\label{eq:64}
    \mathbbm{P}\Big\{\abs{\hat{\beta}_{sk}-\beta_{sk}}\geq \epsilon\Big\}\leq 2e^{-2n\epsilon^2} \quad \forall s,k=1,2.
\end{equation}
Now we can bound $\E(\vert\hat{\beta}-\beta\vert).$ Recall that for any nonnegative random variable $Y,$ $\E(Y)=\int_{0}^{\infty} \mathbbm{P}(Y\geq t) dt.$ Hence, for any $a>0,$
\begin{equation}
\begin{split}
\E(\vert\hat{\beta}-\beta\vert^2)&=\int_{0}^{\infty} \mathbbm{P}(\vert\hat{\beta}-\beta\vert^2\geq t)dt\\
                             &=\int_{0}^{a} \mathbbm{P}(\vert\hat{\beta}-\beta\vert^2\geq t)dt+\int_{a}^{\infty} \mathbbm{P}(\vert\hat{\beta}-\beta\vert^2\geq t)dt\\
                             &\leq a+\int_{a}^{\infty} \mathbbm{P}(\vert\hat{\beta}-\beta\vert^2\geq t)dt.
\end{split}
\end{equation}
Equation (\ref{eq:64}) implies that $\mathbbm{P}(\vert\hat{\beta}-\beta\vert\geq \sqrt{t})\leq 2e^{-2nt}.$ Hence,
\begin{equation}
    \begin{split}
        \E(\vert\hat{\beta}-\beta\vert^2)&\leq a+\int_{a}^{\infty} \mathbbm{P}(\vert\hat{\beta}-\beta\vert^2\geq t)dt\\
                                         &= a+\int_{a}^{\infty}\mathbbm{P}(\vert\hat{\beta}-\beta\vert\geq \sqrt{t})dt\\
                                         &\leq a+2\int_{a}^{\infty}e^{-2nt}dt\\
                                         &=a+\frac{e^{-2na}}{n}.
    \end{split}
\end{equation}
Set $a=\ln(2)/(2n)$ and we have
\begin{equation}
    \E(\vert\hat{\beta}-\beta\vert^2)\leq \frac{\ln(2)}{2n}+\frac{1}{2n}=\frac{1+\ln(2)}{2n}.
\end{equation}
Therefore, we have
\begin{equation}
    \E(\vert\hat{\beta}-\beta\vert)\leq \sqrt{(\E(\vert\hat{\beta}-\beta\vert^2)}\leq \sqrt{\frac{1+\ln(2)}{2n}}.
\end{equation}
Plugging this upper bound back to equation (\ref{eq:47}), we get
\begin{equation}
\begin{split}
    \E\abs{\hat{w}_{ij}-w_{ij}} &\leq \sqrt{\frac{1+\ln(2)}{2n}}(a_iS_i A_{ij}a_jI_j+a_iS_iA_{ij}b_jI_j+ b_iS_iA_{ij}a_jI_j+b_iS_iA_{ij}b_jI_j)\frac{g_i}{\abs{N_i}N}\\
    &(\because \abs{N_i}\geq 1\: \text{and by treating 0 neighbor as equal to}\: 1)\\
    &\leq\sqrt{\frac{1+\ln(2)}{2n}}\frac{A_{ij}g_i}{N}.
\end{split}
\end{equation}
The steps to prove the upper bound for $E_{P^n}\abs{\hat{c}_i-c_i}$ are exactly the same,
\begin{equation}
    \hat{c}_i-c_i = -(\hat{\gamma}_1-\gamma_1)\frac{a_i I_ig_i}{N}-(\hat{\gamma}_2-\gamma_2)\frac{b_i I_ig_i}{N}.
\end{equation}
Take the absolute value and expectation of both sides,
\begin{equation}\label{eq:52}
\begin{split}
    \E\abs{\hat{c}_i-c_i }&=\E\abs{(\hat{\gamma}_1-\gamma_1)\frac{a_i I_ig_i}{N}+(\hat{\gamma}_2-\gamma_2)\frac{b_i I_ig_i}{N}}\\
    &\leq \E\abs{\hat{\gamma}_1-\gamma_1}\frac{a_iI_ig_i}{N}+\E\abs{\hat{\gamma}_2-\gamma_2}\frac{b_iI_ig_i}{N}.
    \end{split}
\end{equation}
With the same idea as for $\beta,$ $\gamma$ is also bounded in $[0,1].$ By using Lemma \ref{lemmahoeff}, we get
\begin{equation}
   \E \abs{\hat{\gamma}-\gamma}\leq \sqrt{\frac{1+\ln(2)}{2n}} \quad \forall\hat{\gamma}=\hat{\gamma}_1,\hat{\gamma}_2.
\end{equation}
Plugging this upper bound back into equation \ref{eq:52}, we get
\begin{equation}
    \begin{split}
      \E \abs{\hat{c}_i-c_i } &\leq \sqrt{\frac{1+\ln(2)}{2n}}(a_i+b_i)\frac{I_ig_i}{N}\\
                             &=\sqrt{\frac{1+\ln(2)}{2n}}\frac{I_ig_i}{N}.
    \end{split}
\end{equation}
\end{proof}
\section{Proofs for Theorems}

\subsection{Proof of Theorem \ref{corollary4.1}}
The objective function $F_n(V)$ is a non-decreasing submodular function for any adjacency matrix, covariate values, and parameter estimates.

\begin{proof}
Recall
\begin{equation}
F_n(V) = \mathbf{v} ^\intercal \hat{W} \mathbf{v} + \hat{C}^\intercal \mathbf{v} - \mathbf{1}_{N\times 1}^\intercal \hat{W} \mathbf{v} - \mathbf{v} ^\intercal \hat{W} \mathbf{1}_{N\times 1}.
\end{equation}
Here, $\mathbf{v}$ is a vector of integers. Let us first, instead, look at this function in the continuous case. Imagine now we have a vector $\Tilde{\mathbf{v}}$ with continuous elements. Then, this function becomes$\colon$
\begin{equation}
\Tilde{F}_n(V) = \Tilde{\mathbf{v}}^\intercal \hat{W} \Tilde{\mathbf{v}} + \hat{C}^\intercal \Tilde{\mathbf{v}} - \mathbf{1}_{N\times 1}^\intercal \hat{W} \Tilde{\mathbf{v}} - \Tilde{\mathbf{v}}^\intercal \hat{W} \mathbf{1}_{N\times 1}. 
\end{equation}
We can write the derivative of $F_n(V)$ with respect to $\Tilde{\mathbf{v}}\colon$
\begin{equation}
\begin{split}
    \frac{\partial \Tilde{F_n}(V)}{\partial \Tilde{\mathbf{v}}}&= \Tilde{\mathbf{v}}^\intercal \hat{W}^\intercal + \Tilde{\mathbf{v}}^\intercal \hat{W}+\hat{C}^\intercal -1_{N\times 1}^\intercal \hat{W}- 1_{N\times 1}^\intercal \hat{W}^\intercal\\
    & = \underbrace{(\Tilde{\mathbf{v}}^\intercal -1_{N\times 1}^\intercal)}_{\leq 0} \underbrace{\hat{W}^\intercal}_{\leq 0}+\underbrace{(\Tilde{\mathbf{v}}^\intercal-1_{N\times 1}^\intercal)}_{\leq 0}\underbrace{\hat{W}}_{\leq 0}+\underbrace{\hat{C}^\intercal}_{\geq 0} \geq 0.
\end{split}
\end{equation}
Given all the elements in $\frac{\partial \Tilde{F}_n(V)}{\partial \Tilde{\mathbf{v}}}$ are non-negative, this non-decreasing property also holds under the integer increment in every element of $\mathbf{v}$.
Therefore, $F_n(V)$ is a non-decreasing set function. Combining this with Lemma \ref{proposition3.1}, we complete the proof.
\end{proof}

\subsection{Proof of Theorem \ref{theorem5.1}}
Let $N_M=\max_{i \in \mathcal{N}} \abs{N_i}$, $N_I$ be the total number of infected units, and $g=\max_{i \in \mathcal{N}} g_i$. Under Assumptions \ref{assumpundir}, \ref{assumpperfect}, and \ref{ass5.1}, we have
\begin{equation}
    \E_{P^n}\Big[F(V^*)-F(\hat{V})\Big]\leq \Bar{C} \cdot \frac{g\big[d\min\{N_M,d\}+2dN_M+\min\{N_I,d\}\big]}{N}\sqrt{\frac{1}{n}}+\frac{1}{e}F(V^*),
\end{equation}
where $\Bar{C}$ is a universal constant and $d$ is the number of available vaccine doses.
\begin{proof}
\begin{equation}\label{eq:theorem5.2}
\begin{split}
    \E_{P^n}\Big[\sup_{V\in\mathcal{V}_d}\abs{F_n(V)-F(V)}\Big]& \leq \E_{P^n}\Big[\sup_{V\in\mathcal{V}_d} \mathbf{v}^\intercal \abs{\hat{W}-W} \mathbf{v}\Big]+\E_{P^n}\Big[\sup_{V\in\mathcal{V}_d} \abs{\hat{C}^\intercal-C^\intercal} \mathbf{v}\Big]\\&\quad+\E_{P^n}\Big[\sup_{V\in\mathcal{V}_d} \mathbf{1}_{N\times 1}^\intercal \abs{\hat{W}-W} \mathbf{v}\Big]+\E_{P^n}\Big[\sup_{V\in\mathcal{V}_d} \mathbf{v}^\intercal \abs{\hat{W}-W} \mathbf{1}_{N\times 1}\Big]\\
    &= \sup_{V\in\mathcal{V}_d}\mathbf{v}^\intercal \E_{P^n}\abs{\hat{W}-W} \mathbf{v}+\sup_{V\in\mathcal{V}_d} \E_{P^n}\abs{\hat{C}^\intercal-C^\intercal} \mathbf{v}\\&\quad+\sup_{V\in\mathcal{V}_d} \mathbf{1}_{N\times 1}^\intercal \E_{P^n}\abs{\hat{W}-W}\mathbf{v}+\sup_{V\in\mathcal{V}_d} \mathbf{v}^\intercal \E_{P^n}\abs{\hat{W}-W} \mathbf{1}_{N\times 1}.
\end{split}
\end{equation}
From equation (\ref{eq:theorem5.2}), $\E_{P^n}\Big[\sup_{V\in\mathcal{V}_d}\abs{F_n(V)-F(V)}\Big]$ can be decomposed into four components. Since $\mathbf{v}$ only contains $\{0,1\}$ and the absolute value must be non-negative, $V$ that maximizes each component under capacity constraint must select units with a greater number of edges, as compare to those that are not selected. We define the maximum number of edges for each unit as $N_M$. Hence, the number of edges for selected units must be lower or equal to $N_M$. Next, we look at each term in equation \ref{eq:theorem5.2} separately. Using Lemma \ref{lamma5.1}, the first term is bounded as$\colon$

\begin{equation}
\begin{split}
    \sup_{V\in\mathcal{V}_d}\mathbf{v}^\intercal \E_{P^n}\abs{\hat{W}-W} \mathbf{v}&\leq
    \sup_{V\in\mathcal{V}_d}\sqrt{\frac{1+\ln(2)}{2n}}\sum_{i\in V}\sum_{j\in V}\frac{A_{ij}g_i}{N}\\
    &\leq \sqrt{\frac{1+\ln(2)}{2n}}\sum_{i\in V}\frac{g_i\min\{N_M,d\}}{N}\\
    &\Big(\because \sum_{j\in V}A_{ij}\leq \min\{N_M,d\} \quad \forall i\in \mathcal{N}\Big)\\
    &\leq \sqrt{\frac{1+\ln(2)}{2n}} \frac{dg\cdot \min\{N_M,d\}}{N}.
    \end{split}
\end{equation}
The second term is bounded as$\colon$
\begin{equation}
\begin{split}
    \sup_{V\in\mathcal{V}_d} \E_{P^n}\abs{\hat{C}^\intercal-C^\intercal} \mathbf{v}&\leq \sup_{V\in\mathcal{V}_d} \sqrt{\frac{1+\ln(2)}{2n}}\sum_{i\in V}\frac{I_ig_i}{N} \leq  \sqrt{\frac{1+\ln(2)}{2n}}\frac{g\min\{N_I,d\}}{N}.
        \end{split}
\end{equation}
The third term is bounded as$\colon$
\begin{equation}
    \begin{split}
    \sup_{V\in\mathcal{V}_d}\mathbf{1}_{N\times1}^\intercal\E_{P^n}\abs{\hat{W}-W}\mathbf{v}&\leq \sup_{V\in\mathcal{V}_d}\sqrt{\frac{1+\ln(2)}{2n}}\sum_{i\in \mathcal{N}}\sum_{j\in V}\frac{A_{ij}g_i}{N}\\
    &\leq \sqrt{\frac{1+\ln(2)}{2n}}\frac{\sum_{j\in V}gN_M}{N}\\
    &= \sqrt{\frac{1+\ln(2)}{2n}}\frac{dgN_M}{N}.
    \end{split}
\end{equation}
The fourth term is bounded as$\colon$
\begin{equation}
    \begin{split}
     \sup_{V\in\mathcal{V}_d} \mathbf{v}^\intercal \E_{P^n}\abs{\hat{W}-W} \mathbf{1}_{N\times 1}   &\leq \sup_{V\in\mathcal{V}_d}\sqrt{\frac{1+\ln(2)}{2n}}\sum_{i\in V}\sum_{j\in \mathcal{N}}\frac{A_{ij}g_i}{N}\\
    &\leq \sqrt{\frac{1+\ln(2)}{2n}}\frac{\sum_{i\in V}gN_M}{N}\\
    &= \sqrt{\frac{1+\ln(2)}{2n}}\frac{dgN_M}{N}.
    \end{split}
\end{equation}
Combining the bounds of the four terms, we get
\begin{equation}
\begin{split}
  \E_{P^n}\Big[\sup_{V\in\mathcal{V}_d}\abs{F_n(V)-F(V)}\Big]&\leq  \sqrt{\frac{1+\ln(2)}{2n}}\frac{dg\min\{N_M,d\}}{N}+ 2\sqrt{\frac{1+\ln(2)}{2n}}\frac{dgN_M}{N}\\&\quad+\sqrt{\frac{1+\ln(2)}{2n}}\frac{g\min\{N_I,d\}}{N}\\
&=\frac{dg\min\{N_M,d\}+2dgN_M+g\min\{N_I,d\}}{N}\sqrt{\frac{1+\ln(2)}{2n}}.
\end{split}
\end{equation}
Therefore, we have from equation (\ref{eq.Vhatregret})
\begin{equation}
\begin{split}
    \E_{P^n}[F(V^*)-F(\hat{V})]&\leq \left(2+\frac{1}{e}\right)\frac{dg\min\{N_M,d\}+2dgN_M+g\min\{N_I,d\}}{N}\sqrt{\frac{1+\ln(2)}{2n}}\\&\quad+\frac{1}{e}F(V^*)\\
    &=\left(2+\frac{1}{e}\right)\frac{g\big[d\min\{N_M,d\}+2dN_M+\min\{N_I,d\}\big]}{N}\sqrt{\frac{1+\ln(2)}{2n}}\\&\quad+\frac{1}{e}F(V^*)
    \end{split}
\end{equation}
Setting $\bar{C} = (2 + 1/e)\sqrt{\frac{1+\ln(2)}{2}}$ completes the proof.
\end{proof}

\end{appendices}
\end{document}